\documentclass[twocolumn,preprintnumbers,amsmath,prc,amssymb]{revtex4}

\usepackage{graphicx}% Include figure files
\usepackage{dcolumn}% Align table columns on decimal point
\usepackage{bm}% bold math

\begin{document}

\title{Critical exponents of the quark-gluon bags  model with the critical endpoint
}

\author{A. I. Ivanytskyi$^1$, K. A. Bugaev$^1$, A. S. Sorin$^2$  and G. M. Zinovjev$^1$}
\affiliation{$^1$Bogolyubov Institute for Theoretical Physics of the
National Academy of Sciences of Ukraine, Metrologichna str. 14$^b$, Kiev-03680, Ukraine
}
\affiliation{$^2$Bogoliubov Laboratory of Theoretical Physics, 
JINR, Joliot-Curie str. 6, 141980 Dubna, Russia
}

%\date{\today}
\begin{abstract}
The critical indices $\alpha'$, $\beta$, $\gamma'$ and $\delta$ of the Quark Gluon Bags with Surface Tension  Model that has  the critical endpoint are calculated and compared with  the exponents  of other models. These indices  are expressed in terms of the most general parameters of the model. Despite the usual expectations the found critical indices do not  depend on the Fisher exponent $\tau$ and  on  the parameter $\varkappa$ which relates the mean bag surface to its volume. The scaling relations for the obtained  critical exponents  are verified and  it is demonstrated that for the standard definition of the index $\alpha'$ the Fisher and the Griffiths scaling inequalities are not fulfilled in general case, whereas the Liberman scaling inequality is always obeyed. 
This is not surprising for the phase diagram with the asymmetric properties of pure phases, but
 the present model  also provides us with the first and  explicit example  that the specially defined index $\alpha'_s$ does not recover  the scaling relations as well. 
Therefore, here we suggest the physically motivated definition of the index $\alpha' = \alpha'_c$ and demonstrate that such a definition recovers the Fisher scaling inequality, while  it is shown that
the Griffiths  inequality should be generalized for the phase diagram with the asymmetric properties.
The critical exponents of several systems that belong to  different universality classes  are successfully described by the parameters of the present model  and hence its equation of state  can be used for a variety of practical applications. 

\vspace*{0.5cm}


\end{abstract}

\maketitle

%%%%%%%%%%%%%%%%%%%%%%%%%%%  Introduction     %%%%%%%%%%%%%%%%%%%%%%%%%%%%%%%%%%%%%%%%%%%%%%%%%%%%%%%%

\section{Introduction}

Scaling has been widely accepted as ``a pillar of modern critical phenomena" \cite{Stanley:99}. 
The scaling hypothesis used in the study of 
critical phenomena was independently developed by the well-known scientists as  Widom, Domb, Hunter, Kadanoff, 
Fisher, Patashinskii and Pokrovskii (see review \cite{Fisher:Rev67} for the  details). 
Further it was developed within the renormalization group approach \cite{Wilson:Rev,RGmethod}.
One of the most striking predictions of the scaling hypothesis is called the scaling laws. 
For instance, for the ordinary liquids these scaling laws  relate the 
critical exponents $\alpha$, $\beta$, $\gamma$ and $\delta$ which describe the behavior of the specific heat capacity 
($C \sim |t|^{- \alpha^\prime} $, here $t = \frac{T-T_ {cep}}{T_ {cep}} < 0$ denotes a relative deviation of the temperature $T$ from the critical one $T_{cep}$), density differences of the liquid and gaseous  phases  ($\rho_l -\rho_g \sim |t|^{ \beta} $),  isothermal compressibility ($K_T\sim |t|^{-\gamma^\prime}$)  and the shape of the critical isotherm which  is given by the critical index $\delta$ (for the formal definitions see below)
\begin{eqnarray}
\label{F1}
{\rm Fisher ~ [5]: %\cite{Fi:64}:
} & \hspace{.5cm}	\alpha^\prime + 2\beta + \gamma^\prime ~& \ge~2 ~,\\
\label{G}
{\rm Griffiths ~ [6]:%\cite{Gri:65}:
}& 		\alpha^\prime + \beta(1 + \delta) & \ge~ 2 ~,\\
\label{L}
{\rm Liberman ~ [7]:%\cite{Li:66}:
}&			\gamma^\prime + \beta(1-\delta) & \ge ~0 ~.
\end{eqnarray}
Similar equalities can be also introduced for  magnetic systems \cite{Stanley:71} and for  percolating systems \cite{Percolation}.  The corresponding exponent inequalities for magnetic systems are often called Rushbrooke's \cite{Rushbrooke}, Griffiths' and Widom's \cite{Widom} inequalities, respectively. 

The superscript prime in Eqs. (\ref{F1})--(\ref{L}) is necessary to introduce for the  systems with the  phase diagram  for   which the behavior of specific heat capacity and/or compressibility on the gaseous  side of phase diagram, where $t < 0$, differs from that one on the liquid side, i.e. for  $t > 0$. 
Although in many physical systems the scaling laws (\ref{F1})--(\ref{L}) are obeyed as equalities, it is customary to write them as inequalities since, as was proven by M. E. Fisher 
 for liquids \cite{Fi:64,Fi:69}, in the most general case they can be established as inequalities only. 
It was, however, found that in such  exactly solvable models  
as the Fisher-Felderhof one-dimensional model 
\cite{Fi:70}, the  statistical multifragmentation model (SMM) \cite{Bugaev_00}, the quark gluon bags with surface tension model (QGBSTM1) with the  tricritical endpoint \cite{Bugaev_07_physrev} and its generalization \cite{FWM},   the Fisher   and  the Griffiths scaling inequalities  (\ref{F1}) and (\ref{G}), 
in which  the index $\alpha^\prime$ is involved,  may be broken. 
The corresponding proofs 
are given in \cite{Fi:70} for the Fisher-Felderhof  model, in \cite{Reuter_01} for the SMM 
and in \cite{Ivanytskyi} for  the QGBSTM1 and its generalization \cite{FWM}. 
In all these cases the phase diagram is rather asymmetric, i.e. the behavior  of pure phases on the both sides of the phase equilibrium curve are different, 
 and, as a result,  the heat capacity of  gaseous and liquid phases are quite  different. 
To some extent this problem was resolved  by M. E. Fisher in \cite{Fi:70}  by  introduction of a specially defined index $\alpha_s^\prime$
which measures a divergency of the heat capacity  difference of two phases at the critical endpoint (CEP). In this case the $\alpha_s^\prime$ index gives the maximal 
 value among  the $\alpha^\prime$ index  of the gaseous and the liquid phases. 
The usage of such an index instead of the traditional exponent $\alpha^\prime$ allows one to formally recover  the Fisher and the Griffiths inequalities in  all  exactly solvable models mentioned above,  although neither the physical meaning of the index $\alpha_s^\prime$ nor its relation to the experimental procedure of the heat capacity  measurement were  ever justified. 
Therefore, we are faced to three  principal questions: (I) Is it possible to justify  the $\alpha_s^\prime$  definition?  (II) What definition should be used instead of  the $\alpha_s^\prime$ index, if it fails to recover the scaling inequalities  (\ref{F1})--(\ref{L})? (III) 
What should we do with  the scaling laws  in the latter case? 

In order to clarify  these questions  in the present work  we calculate   the critical exponents of  the quark gluon bags with surface tension model  (QGBSTM2 hereafter) which has  the CEP \cite{Bugaev_09}. 
This  phenomenological model is a novel development of the well known gas of bags model \cite{GasOfBags:81}
which, however,  contains  entirely  new  mathematical  mechanism of the CEP  generation.
Of course, the  QGBSTM2 was developed for the CEP modeling without  specifying  its universality class and the present work is devoted to the finding of the critical exponents of this model and to the
determination of its universality classes. 
Since our main subject of interest is to describe the endpoint  properties of the deconfinement   phase transition (PT)  of quantum chromodynamics (QCD), we would like to pay a special attention 
to the QCD CEP properties. For this purpose  we, in accord with the contemporary knowledge,   suppose that the CEP of   the deconfinement  PT has the  properties typical for 3-D Ising  model in the case 
of  3 quark flavors  degenerated QCD   \cite{IsingQCD1,IsingQCD2,IsingQCD3}, whereas for 2+1 quark flavors we assume that  the QCD endpoint belongs to  the  universality class of the O(4) symmetric 3-dimensional spin model
\cite{Karsch:01,Rob:84,Wilczek:92,Ejiri:09,Kaczmarek:11}. In other words  we would like to determine the QGBSTM2 parameters and fix them  in order to reproduce 
the critical exponents  of   respective  universality class.

In contrast to the comparable solvable  models \cite{Fi:70,Bugaev_00,Bugaev_07_physrev, FWM} including the QGBSTM1, the QGBSTM2 has entirely 
different structure of isobaric  ensemble  singularities describing the PT, and, as we show below, for some values of parameters   the Fisher and the Griffiths inequalities 
(\ref{F1}) and (\ref{G})  of this model  are not fulfilled for  both  $\alpha^\prime$ and $\alpha_s^\prime$ indices.  
Therefore, here we introduce 
 a physically motivated definition of the supremum  index $\alpha_c^\prime$  which is    found from  the linear combination of the specific  heat capacities of pure phases taken with the nonsingular weights.  
On the examples of solvable models \cite{Fi:70,Bugaev_00,Bugaev_07_physrev, FWM}  we demonstrate  that such a definition recovers the  scaling inequalities in those case, when  the traditional $\alpha^\prime$ index fails, since  $\alpha_c^\prime \ge \alpha_s^\prime$. 
 However,  below we show that even with an improved  definition  of the index $\alpha^\prime = \alpha_c^\prime$ for  the QGBSTM2 critical exponents  there are two regimes:
 the traditional scaling regime, when 
the scaling  inequalities (\ref{F1})--(\ref{L}) are   held  as the equalities only, and 
the generalized scaling regime, when the Fisher (\ref{F1}) and the Liberman (\ref{L}) inequalities are fulfilled, but the Griffiths inequality (\ref{G}) is only obeyed in its generalized form which is suggested here. 
The performed thorough  analysis of the scaling laws (\ref{F1})--(\ref{L}) and their generalization onto the case  of   the phase diagram with the asymmetric  properties  seem  to be  very important nowadays in a view of fast technological and computational  progress which, respectively,  allows one  to  study 
 the substances and  models with new and unusual thermodynamic properties.

The work is organized as follows. Section \ref{secmodel} is devoted to a brief discussion of the QGBSTM2 main ingredients. The model is analyzed in details in Section \ref{seccritical indices}. The QGBSTM2 critical exponents are calculated in that section also. The analysis of the scaling relations between the found critical exponents is given in Section \ref{secscallings}. Conclusions and perspectives are discussed in Section \ref{secconclusions}.

%%%%%%%%%%%%%%%%%%%%%%%%%%%   Model   %%%%%%%%%%%%%%%%%%%%%%%%%%%%%%%%%%%%%%%%%%%%%%%%%%%%%%%%%%%%%%%%%%%%%%%%%%%%%%%%

%
\section{QUARK GLUON BAGS WITH SURFACE TENSION MODEL}
\label{secmodel}

The QGBSTM2  \cite{Bugaev_07_physrev,Bugaev_09}  treats the  quark-gluon plasma (QGP) bags and hadrons as relevant degrees of freedom.
Similarly to the original statistical bootstrap model \cite{SBM} 
the attraction between the degrees of freedom in this model is accounted via many sorts of the constituents, while 
the repulsion between them  is introduced a la Van der Waals equation of state \cite{GasOfBags:81,Bugaev_07_physrev,Bugaev_09}. 
The phase structure of the QGBSTM2 is completely defined by the mass-volume spectrum
that for a given temperature $T$, baryonic chemical potential $\mu$ is defined as  
\begin{eqnarray}
\label{EqIV}
F (z,T,\mu) & = & F_H(z,T,\mu)+u(T,\mu)I_\tau(\Delta z,\Sigma)  \,.
\end{eqnarray}
This spectrum defines   the isobaric partition \cite{GasOfBags:81,Bugaev_07_physrev,Bugaev_09}
\begin{eqnarray}
\label{EqV}
{\cal Z} (z,T,\mu)  = \frac{1}{z~ -~F (z,T,\mu) }\,,
\end{eqnarray}
where $z$ denotes the isobar variable.  

The  discrete part of the mass-volume spectrum $F_H$ in (\ref{EqIV}) is 
successfully used as the hadron resonance gas model to describe the experimental hadron multiplicities 
which allow one to recover the thermodynamic quantities of strongly interacting matter created in the heavy ion collisions, when this matter reaches  the  chemical freeze-out stage (an incomplete list of related works can be found in \cite{Andronic:05,Oliinychenko:12}). 
Here we consider the simplest parameterization of  $F_H$, since both the quantum statistics and the width of hadron resonances are important for the  temperatures below 50 MeV and the baryonic chemical  potentials  larger than  940 MeV  \cite{Andronic:05,Oliinychenko:12}. 
Thus the spectrum $F_H$ is  parameterized as follows
\begin{eqnarray}\label{EqVI}
F_H(z,T,\mu) & = & \sum_{j=1}^n g_j e^{\frac{b_j \mu}{T} -v_j  z} \phi(T,m_j) \,. 
\end{eqnarray}
The continuous part of the spectrum (\ref{EqIV}) is chosen in the simplest form (compare it that one used in \cite{FWM}), which can be  cast as  an  integral 
($V_0 \approx 1$ fm$^3$ \cite{Shelest:82})
\begin{eqnarray}
\label{EqVII}
I_\tau(\Delta z,\Sigma) & = & \int\limits_{V_0}^{\infty}\frac{dv}{v^\tau}e^{-\Delta z \, v-\Sigma v^{\varkappa}} \,,
\end{eqnarray}
where the Fisher exponent $\tau > 2$ provides the convergence of the integral (\ref{EqVII}) for $\Delta z = 0$ and $\Sigma = 0$. 
Also here the notation  $\Delta z\equiv z -z_M(T,\mu)$ is introduced.  In (\ref{EqVI}) the particle density of a  hadron of mass $m_j$, baryonic charge $b_j$, eigenvolume $v_j$ and degeneracy $g_j$ is denoted as
\begin{equation}
\phi(T,m_j)\equiv \frac{1}{2\pi^2}\int\limits_0^{\infty}\hspace*{-0.1cm}p^2dp\,
e^{\textstyle-\frac{(p^2~+~m_j^2)^{1/2}}{T}}.
\end{equation}
The  functions $u(T,\mu)$ and $z_M(T,\mu)$ in (\ref{EqIV}) and (\ref{EqVII}) are the parameters of the present model which are smooth and  finite together with all  their  first and second derivatives \cite{Bugaev_07_physrev,Bugaev_09}. 
The $z$-dependent exponentials  in (\ref{EqVI}) and (\ref{EqVII}) describe the short range repulsion of the Van der Waals type \cite{Bugaev_07_physrev,Bugaev_09}.   To parameterize the surface of a QGP bag in the continuous  part of the spectrum  (\ref{EqVII}) the parameter $\varkappa$ is introduced. Usually the constant $\varkappa$ is defined by the dimension $d$ as $\varkappa=\frac{d-1}{d}$, but in what follows  it is treated as a free parameter with   the range of values  $0<\varkappa<1$.

A few words should be added  here about   the hadronic surface tension. In principle, it can be included into the QGBSTM2 discrete spectrum (\ref{EqVI}). 
The first and interesting results about the surface tension of hadrons   which fit well into 
the QGBSTM2 framework  can be found in  \cite{Oliinychenko:12}. They clearly demonstrate that the hadronic surface tension is rather small, although it changes the sign at the temperature about 150 MeV. However, 
in the present work we do not consider this element  and set the hadronic surface tension to zero since 
its inclusion  does not affect the expressions  for the critical indices.
 
The new element of principal importance of the present model is the parameterization of the  surface tension coefficient $T \Sigma(T,\mu)$ ($\Sigma(T,\mu)$ in (\ref{EqVII})  denotes  the reduced surface tension coefficient) which in the vicinity of the phase equilibrium curve $T= T_c(\mu)$   is defined as 
\begin{eqnarray}
\label{EqIX}
\Sigma^\pm(T,\mu)&=&\mp\frac{\sigma_0}{T}\cdot
%%%\left( T_{cep}-T_c(\mu)+\frac{dT_c}{d\mu}(\mu_{cep}-\mu)\right)^{\xi^\pm}
\left( T-T_c(\mu_{cep})+\frac{dT_c}{d\mu}(\mu_{cep}-\mu)\right)^{\xi^\pm}
\nonumber \\
&\times&\left| \frac{T_\Sigma(\mu)-T}{T_\Sigma(\mu)}\right|^{\zeta^\pm} \,,
\end{eqnarray}
with the following values of constants $\zeta^\pm\ge1$, $\xi^\pm>0$. Here $\sigma_0$ is chosen to be a positive constant, but  the obtained results hold, if $\sigma_0 >0$ is a smooth function of $T$ and $\mu$. 
As shown below it is also of crucial importance that the parameters $\zeta^\pm$ and $\xi^\pm$  have different values below and above the phase coexistence curve $T = T_c(\mu)$ which exists for $\mu \ge \mu_{cep}$. 
It can be shown \cite{Bugaev_09} that  the necessary condition for the deconfinement PT existence with the  CEP    is that    the QGBSTM2 surface tension coefficient changes the sign exactly at the phase equilibrium curve.
In other words, the solution $T_\Sigma(\mu)$  of the equation $\Sigma(T,\mu) = 0$ should coincide 
with  the PT curve $T_c(\mu)$ for $\mu \ge \mu_{cep}$, i.e. 
 $T_\Sigma(\mu)=T_c(\mu)$ for $\mu \ge \mu_{cep}$ (see Fig. 1 for details).  
The important physical consequence of the nil surface tension  curve and the deconfinement PT curve matching  is that for  the corresponding thermodynamic parameters the volume (and, hence, the mass) distribution of  QGP  bags becomes the power-like instead of exponential. Typically,
 in the liquid-gas PT  the power-like volume distribution of droplets  corresponds to the CEP \cite{Fisher_67} and it leads to the formation of  fractals.  Such a power law may, in principle, naturally  explain  the appearance of  the  non-Boltzmann fluctuations in the high energy collisions experiments \cite{Phenix:Fluc,CMS:Fluc} without appealing to the Tsallis statistics \cite{Wilk:Fluc10}. 
 
The important mathematical consequence of such a matching is that 
the discontinuity of  
 the partial $\mu$ and $T$ derivatives  of the reduced surface tension coefficient  across the line $T= T_\Sigma(\mu)$ provides the 1st order deconfinement PT existence  \cite{Bugaev_09}, and hence  in  (\ref{EqIX}) one has $\zeta^+\neq\zeta^-$ and $\xi^+\neq\xi^-$ in general.
The quantities introduced in (\ref{EqIX}) have the superscript $+(-)$, if they are taken for $T$ above (below) the curve 
$T_\Sigma(\mu)$ in the whole $\mu-T$ plane. 
For $0 \le \mu <  \mu_{cep}$ the nil line of the surface tension coefficient  is located in such a way that  the deconfinement PT degenerates into a cross-over since in this region $\Sigma(\mu)< 0$ (see \cite{Bugaev_09} and  Fig. 1 for more details).
  
Also note that  the different slopes of the surface tension coefficient below and above its nil line
$T= T_\Sigma(\mu)$ are not unusual since this  property is successfully  used in such well known models as the 
Fisher droplet model (FDM) \cite{Fisher_67} and  the SMM \cite{Bondorf,Bugaev_00}, but, additionally,  in the present model 
the reduced surface tension coefficient  is negative (positive) for $T$ above (below) the  line $T= T_\Sigma(\mu)$. As it is argued  in \cite{Bugaev_07_physrev,Bugaev_09} there is nothing wrong or unphysical with the negative values of surface tension coefficient, since in the grand canonical ensemble the quantity $T \Sigma\, v^\varkappa$ is the surface  free energy $f_{surf}=e_{surf}-Ts_{surf}$ of the bag of mean volume $v$, were $e_{surf}$ and $s_{surf}$ are the surface energy and entropy. Therefore, $\Sigma<0$ means  that the surface entropy contribution simply exceeds the surface energy part, i.e. $Ts_{surf} > e_{surf}$ and then $f_{surf}<0$.
It can be shown on the basis of exactly solvable model of surface deformations  \cite{Bugaev_05_Physrev} that  negative values of the surface free energy  is a consequence  of very large  number of non-spherical configurations at high temperatures. To our best knowledge, 
the exactly solvable models of  the liquid-gas PT 
with  negative values of  the surface tension coefficient  provide us with  the only physical  reason preventing the condensation of small droplets into a liquid phase (an infinite droplet) at supercritical temperatures, and, thus, they  naturally explain  the existence of a cross-over  both in QGP   \cite{Bugaev_07_physrev,Bugaev_09} and, probably, in the ordinary liquids \cite{Bugaev_11}.
For the field-theoretical arguments  in favor of the negative surface tension of quark gluon bags see \cite{Hosek:91,Hosek:93}. 
Another strong line of  arguments  in favor of the negative surface tension of quark gluon bags  at high temperatures is provided by the  recent analysis of  the relation between the confining  color string tension and the surface tension of QGP bag \cite{String:09}. It clearly demonstrates that at the cross-over region the surface tension coefficient of  large bags  is unavoidably negative and, as shown  in \cite{String:09}, this should lead
 to an appearance of  surfaces with the fractal dimension.  Note that the fractal surfaces  are well known in the lattice QCD formulation  \cite{Fractals:QCD1, Fractals:QCD2}, but their principal role in  the
 lattice entropy  maximum  formation  of  the confining  tube   (the so called `mysterious maximum'  \cite{Shuryak:sQGP})  and their  relation to the negative surface tension values of such tubes   were  revealed only recently \cite{String:09}.
  
The phases of the QGBSTM2 include the hadronic phase and the QGP which in the $\mu-T$ plane are separated 
by the nil line of the surface tension coefficient  and they can be distinguished by the sign of the surface tension coefficient (see Fig. 1). For a given $\mu$  and temperature $T$  above (below)  the line $T_\Sigma(\mu)$  the  pressure  is marked by the superscript $+(-)$ and is defined by the equation 
\begin{eqnarray}
\label{EqX}
p^\pm(T,\mu)& = & T\left[F_H(z^\pm,T,\mu)+u(T,\mu)I_\tau(\Delta^\pm z,\Sigma^\pm)\right] \,\hspace*{-0.1cm}  ,\hspace*{0.6cm}  
\end{eqnarray}
where the following notations $z^\pm\equiv\frac{p^\pm(T,\mu)}{T}$, $\Delta^\pm z\equiv z^\pm (T,\mu)-z_M(T,\mu)$ are used. The expression (\ref{EqX}) for pressures $p^+$ and $p^-$ is determined, respectively,  by  the simple  poles $s^+$ and $s^-$ of the isobaric partition (\ref{EqV}).

The  mixed phase of these two phases corresponds to the vanishing value of the surface tension and in this respect it is similar to the CEP in ordinary liquids \cite{Stanley:71}, but,  in contrast to the ordinary liquids, the  PT in the QGBSTM2 is not of the 2-nd order, but of the 1-st order everywhere at the phase diagram except for  the CEP where it is, indeed, of the 2-nd order \cite{Bugaev_09}. The corresponding pressure is given by the essential singularity of the isobaric partition (\ref{EqV})
\begin{eqnarray}
\label{EqXI}
p_M(T,\mu) & = & Tz_M(T,\mu) \,. 
\end{eqnarray}
This  equation also demonstrates the meaning of the function $Tz_M(T,\mu)$ which would give the QGP pressure in the absence of the surface tension. 

As usual, the pressure of the stable phase is defined by the rightmost singularity of the isobar partition (\ref{EqV}) \cite{Fi:70,Bugaev_00,Bugaev_07_physrev,FWM, Reuter_01,Ivanytskyi,Bugaev_09}.  By construction  the PT occurs at $T = T_\Sigma(\mu)$ at which  the simple pole singularities $z^\pm (T,\mu)$  coincide with the essential singularity  $z_M (T,\mu)$. The colliding singularities automatically provide the  fulfillment 
of the Gibbs criterion of phase equilibrium.  The necessary condition for  the PT existence is the following relation between the parameters of the model 
\begin{eqnarray}
\label{EqXII}
z_M(T,\mu) & = & F_H(z_M (T,\mu),T,\mu)+u(T,\mu)I_\tau(0, 0)\,, ~~
\end{eqnarray}
at $T = T_c (\mu) \equiv  T_\Sigma(\mu)$.

\begin{figure}[ht]
\centerline{\includegraphics[width=8.5cm, height=5cm]{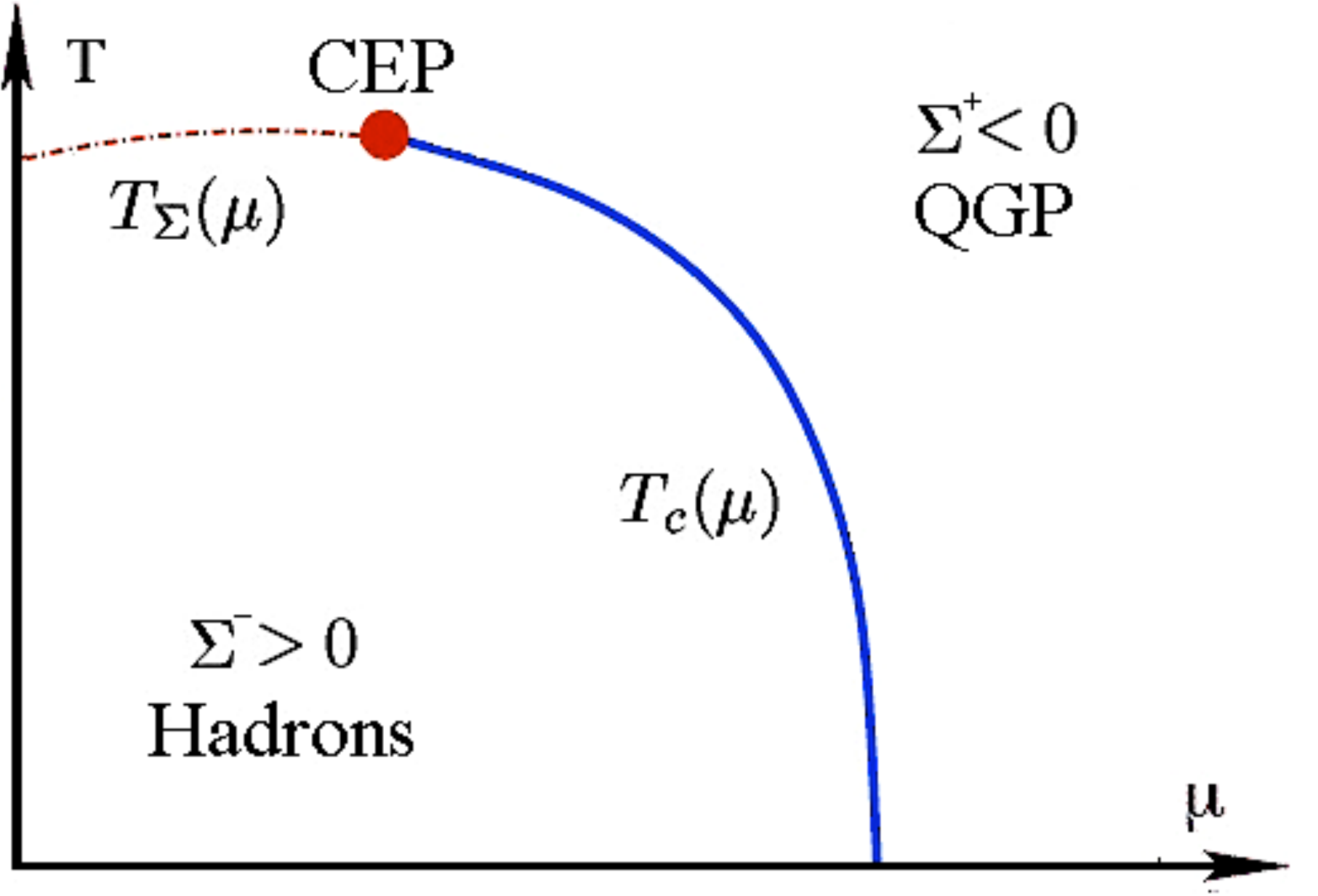}}
\vspace*{0.3cm}
\caption{[Color online]
A schematic  phase diagram in the plane of baryonic chemical potential $\mu$ and temperature $T$. The  deconfinement PT line  $T_c(\mu)$ is shown by the full curve  for $\mu > \mu_{cep}$, whereas the cross-over (shown by dashed curve) takes place along the line $T_\Sigma(\mu)$ for $\mu < \mu_{cep}$. For $\mu \ge  \mu_{cep}$ the reduced  surface tension coefficient $\Sigma$ changes the sign exactly at the PT line, i.e.  $T_c(\mu)=T_\Sigma(\mu)~{\rm for}~\mu \ge \mu_{cep}$. The   PT region ends at  the CEP (filled  circle).
}
\label{fig_T-Mu}
\end{figure}
Assuming that (\ref{EqXII}) is fulfilled we parameterize the shape of the phase coexistence curve $T_c(\mu)$ 
in the vicinity of the CEP  by the  constant  $\xi^T>0$:
\begin{equation}
\label{EqXIII}
T_{cep}-T_c(\mu)\sim(\mu-\mu_{cep})^{\xi^T} \,.
\end{equation}
The crucial  importance of this  new index   for the QGBSTM1 was  demonstrated recently
 \cite{Ivanytskyi}. Surprisingly this  index was not considered  in such  well known models as  the FDM  \cite{Fisher_67} and the SMM \cite{Bondorf,Bugaev_00},
although it is known that  a related quantity $K_c$ introduced in \cite{Soldat:99} plays a decisive role in the classification of the CEP  stability types \cite{Soldat:05}.
Below it will be shown that  just  this index determines the values of the  exponent $\alpha'$.

Using the standard  definitions for  the entropy density  $s$ and the baryonic density $\rho$ as $T$ and $\mu$ partial derivatives of the corresponding pressure   one  can explicitly write the Clapeyron-Clausius equation for pure  phases $\frac{d\mu_c}{dT}=-\frac{s^+-s^-}{\rho^+-\rho^-}\bigl|_{T=T_c}$ with the help of (\ref{EqX}). However, in the present model there is an additional relation for the pressure of the mixed phase $p_M$ (\ref{EqXI}), which by construction  matches the pressures $p^-$ and $p^+$ (\ref{EqX})  for the same value of the  PT temperature. Therefore, one can establish two additional relations of the Clapeyron-Clausius type between the partial derivatives of the function $p_M(T, \mu)$ and the partial derivatives of the pressure of each pure phase 
\begin{equation}\label{EqXIV}
\frac{d\mu_c}{dT}=-\frac{s_M-s^\pm}{\rho_M-\rho^\pm}\biggl|_{T=T_c}=-\frac{A_T}{A_\mu}\biggl|_{T=T_c},
\end{equation}
where the following notation
\begin{equation}\label{EqXV}
A_x=\frac{\partial F_H}{\partial x}+\frac{\partial u}{\partial x}I_\tau+
\frac{\partial z_M}{\partial x}\left(\frac{\partial F_H}{\partial z}-1\right)
\end{equation}
for $x \in \{T,\mu\}$ is used.

For further evaluation  it is  convenient to  parameterize the  behavior of  the numerator and denominator in (\ref{EqXIV}) at the CEP vicinity in the same way as it was suggested recently  in \cite{Ivanytskyi}:
\begin{eqnarray}\label{EqXVI}
A_T\bigl|_{T=T_c} & \sim & (T_{cep}-T_c(\mu))^{\chi+\frac{1}{\xi^T}-1},
\\
\label{EqXVII}
A_\mu\bigl|_{T=T_c} & \sim & (T_{cep}-T_c(\mu) )^\chi,
\end{eqnarray}
where  the finite values of the integral $I_\tau (0,0)$ and the functions 
$F_H (z^\pm (T,\mu),T,\mu)$, $u (T,\mu)$, $z_M (T,\mu)$ along  with their first  derivatives for any finite values of  $T$ and $\mu$ provide the validity of the  condition $\chi\ge\max(0,1-\frac{1}{\xi^T})$. Since the index $\chi$  unavoidably  appears from an inspection of the  Clapeyron-Clausius equation, which is a direct consequence of the Gibbs criterion, then it is quite general \cite{Ivanytskyi}. 
Moreover, the parameter $\chi$ played an important role in  separating the different sets of solutions  for the QGBSTM1 \cite{Ivanytskyi}, but as it will be shown below, although such 
an index appears in the intermediate expressions for the analyzed singularities, surprisingly,  it does not enter any equation for the critical exponents of the present model.

%%%%%%%%%%%%%%%%%%%%%%%%%%%% Critical indices  %%%%%%%%%%%%%%%%%%%%%%%%%%%%%%%%%%%%%%%%%%%%%%%%%%%%%%%%%%%%%%%%%%%%

%
\section{THE STANDARD  CRITICAL INDICES OF THE QGBSTM2}
\label{seccritical indices}

As usual, 
the standard set of the  critical exponents  $\alpha'$, $\beta$ and $\gamma'$ \cite{Fi:64, Fi:70, Stanley:71} describes the $T$-dependence of the system near the CEP:
\begin{eqnarray}
\label{EqXVIII}
C_\rho  & \sim\ &
|t|^{-\alpha'},  \hspace*{0.1cm}
{\rm for}  \quad  t \le 0\quad{\rm and}\quad\rho=\rho_{cep},\\
\Delta\rho & \sim  & |t|^\beta,  \hspace*{0.4cm}
{\rm for}  \quad t \le 0, \\
\label{EqXX}
\Delta K_T & \sim  & |t|^{-\gamma'},  \hspace*{0.1cm}
{\rm for}  \quad t < 0,
\end{eqnarray}
where $\Delta\rho\equiv(\rho^+-\rho^-)_{T=T_c}$ defines the order parameter, $C_\rho\equiv\frac{T}{\rho}(\frac{\partial S}{\partial T})_\rho$ denotes the specific heat capacity at the critical density and $\Delta K_T\equiv(K_T^--K_T^+)_{T=T_c}$ is the discontinuity in the isothermal compressibility $K_T\equiv\frac{1}{\rho}(\frac{\partial\rho}{\partial p})_T$ across the PT line, the variable $t$ is the  reduced temperature $t \equiv \frac{T-T_{cep}}{T_{cep}}$.
The   critical isotherm shape is given by the  index $\delta$ \cite{Fi:64,Stanley:71}
\begin{equation}
p_{cep}-\widetilde{p}\sim(\rho_{cep}-\widetilde{\rho})^\delta  \hspace*{0.5cm}
{\rm for}  \quad t = 0.
\end{equation}
Hereafter the tilde indicates that $T=T_{cep}$. Note that for the phase diagram shown in Fig. \ref{fig_T-Mu} one has  $\widetilde{p}=p^+$ and $\widetilde{\rho}=\rho^+$.

%%%%%%%%%%%%%% Alpha' %%%%%%%%%%%%%%%%%%%%%%%%%%%%%%%%%%%%%%%%%%%%%%%%%%%%%%%%%%%%%%%%

The critical exponent $\alpha'$ describes the $T$-behavior of the specific heat capacity along the critical isochore  $\rho=\rho_{cep}$ inside the mixture of the  QGP and hadron phase. 
Similarly to  \cite{Ivanytskyi} here we do not use the  Yang-Yang formula \cite{YangYang:64}   to calculate $C_\rho$, since it leads to a less convenient representation. Therefore,   some details of the  heat capacity evaluation are given below.
 We have to note that there exist a few suggestions  on how to possibly determine the specific heat capacity at  the vicinity of  the QCD CEP \cite{Misha:98,Nonaka}  from the experimental data obtained in the heavy ion collisions and, hence, the  question of correct definition of  the $\alpha'$ index becomes of crucial importance for us.

As usual, the entropy density  and the  baryonic density of the mixed phase are defined via the corresponding values of  pure phases and  the  volume fraction of hadronic phase  $\lambda$ (the volume fraction of the QGP is, respectively, $1-\lambda$):
\begin{eqnarray}
\label{EqXXII}
\rho|_{T=T_c} & = & \lambda\rho^-|_{T=T_c}+(1-\lambda)\rho^+|_{T=T_c}, \\
\label{EqXXIII}
s |_{T=T_c} & = & \lambda s^-|_{T=T_c}+(1-\lambda)s^+|_{T=T_c}.
\end{eqnarray}
Varying  $\lambda$ from 0 to 1 one can describe all states inside  the mixed phase. Fixing  $\rho|_{T=T_c}$ to  $\rho_{cep}$  in  (\ref{EqXXII}) one can first calculate the  total $T$-derivative of  the volume fraction $\lambda$ along the critical isochore $\rho=\rho_{cep}$
and then one  can  use (\ref{EqXXIII}) to determine the specific heat capacity at fixed baryonic density 
\begin{eqnarray}
C_{\rho}&=&\frac{T}{\rho_{cep}}\left[\frac{ds^+}{dT}+\frac{d\mu_c}{dT}\frac{d\rho^+}{dT}\right.
\nonumber\\
&+&\left.
\lambda\left(\frac{d}{dT}(s^--s^+)+\frac{d\mu_c}{dT}\frac{d}{dT}(\rho^--\rho^+)\right)\right]_{T=T_c}
\nonumber \\
\label{EqXXIV}
&=&\frac{T}{\rho_{cep}}\left[\frac{ds^+}{dT}+\frac{d\mu_c}{dT}\frac{d\rho^+}{dT}
+(\rho^+-\rho_{cep})\frac{d^2\mu_c}{dT^2}\right]_{T=T_c},\hspace*{-0.4cm},\hspace*{0.6cm}
\end{eqnarray}
where in the second step the Clapeyron-Clausius equation (\ref{EqXIV})  for hadron phase  and that one for   the QGP together with the definition   (\ref{EqXXII}) for $\lambda$ were used. 
Using    the Clapeyron-Clausius equation (\ref{EqXIV}) for the QGP 
one can again rewrite (\ref{EqXXIV}) as 
\begin{eqnarray}
\hspace*{-0.1cm}
C_{\rho}\hspace*{-0.1cm}&=&\hspace*{-0.2cm}
\frac{T}{\rho_{cep}}\left[\frac{ds_M}{dT}+\frac{d\mu_c}{dT}\frac{d\rho_M}{dT}+\frac{d}{dT}(s^+-s_M)\right.
\nonumber\\
&+&\hspace*{-0.1cm}
\left.\frac{d\mu_c}{dT}\frac{d}{dT}(\rho^+-\rho_M)+(\rho^+-\rho_{cep})\frac{d^2\mu_c}{dT^2}\right]_{T=T_c} \nonumber \\
\label{EqXXV}
&=&\hspace*{-0.1cm}
\frac{T}{\rho_{cep}}\left[(\rho_M-\rho_{cep})\frac{d^2\mu_c}{dT^2}+
\frac{ds_M}{dT}+\frac{d\mu_c}{dT}\frac{d\rho_M}{dT}\right]_{T=T_c} \,\hspace*{-0.3cm} . \hspace*{0.6cm}
\end{eqnarray}
The obtained expression (\ref{EqXXV}) for the specific heat capacity coincides with  that one obtained for the  QGBSTM1 \cite{Ivanytskyi} and, therefore, the  $\alpha'$-index  for the models  with  the CEP and with the  triCEP are identical.

In this section we consider the divergent heat capacity  at the CEP, i.e. $\alpha' >0$,  while the 
negative values of this index are analyzed in the subsequent section. 
Then the critical exponent $\alpha' > 0$ describes the  temperature behavior of the most singular term in Eq. (\ref{EqXXV}).  Expanding  $\rho_M|_{T=T_c}$ into the series of $t$-powers  and using the  parameterization (\ref{EqXIII})  of  the PT  curve one can show  that the  first term  staying 
on the right hand side of 
 (\ref{EqXXV}) after the second equality  behaves as $t^{\min(1,\frac{1}{\xi^T})+\frac{1}{\xi^T}-2}$. Since the entropy density and the baryonic  density of QGP are, respectively,   the $T$ and $\mu$ derivatives of the function $Tz_M(T,\mu)$, which is a  regular function of its parameters  together   with its first and second  derivatives,   then  the singularity in  the second and third terms of (\ref{EqXXV})  may appear  from 
the square of the derivative $\frac{d\mu_c}{dT}$ and it has the form 
$\left(\frac{d\mu_c}{dT}\right)^2 \sim t^{\frac{2}{\xi^T}-2}$.  Accounting for these facts,  one gets the critical exponent $\alpha'$  as
\begin{equation}\label{EqXXVI}
\alpha'=2-2\min\left(1,\frac{1}{\xi^T}\right).
\end{equation}
This equation shows that $\alpha'>0$ for $\xi^T>1$ only,  otherwise  $ \alpha'=0$. 
As it was mentioned  above, the critical exponent $\alpha'$ of the QGBSTM2 (\ref{EqXXVI})  exactly coincides with that one obtained for the QGBSTM1 \cite{Ivanytskyi}
despite the fact that the phase diagrams of these models have  essential differences. Moreover, these models have entirely different  ranges of the Fisher exponent $\tau$: 
$\tau>2$ within the QGBSTM2 while $1<\tau\le2$ within the QGBSTM1.

%%%%%%%%%%%%%%%%%%% Series Itau %%%%%%%%%%%%%%%%%%%%%%%%%%%%%%%%%%%%%%%%%%%%%%%%%%%%%%%%%

To calculate the critical exponents $\beta$ and $\gamma'$ we need to know the behavior of the integrals $I_{\tau-q}(0,0)$. As it was shown in \cite{Bugaev_09} the necessary condition for the 1-st order deconfinement PT existence is that  such  integral should be finite for $q=1$. This condition provides the range of  values $\tau>2$ for the Fisher exponent.
The validity of this necessary condition can be demonstrated as follows.
Indeed, using the 
 definition of the baryonic density and  Eqs. (\ref{EqX}) for the phases on two sides of the PT curve   one can  find the baryonic density discontinuity across the deconfinement PT line \cite{Bugaev_09}:
\begin{eqnarray}
\label{EqXXVII}
\Delta\rho &=&   -T\cdot\frac
{\left(\frac{\partial\Sigma^+}{\partial\mu}-\frac{\partial\Sigma^-}{\partial\mu}\right)uI_{\tau-\varkappa}(0,0)}
{1-\frac{\partial F_H}{\partial z}+uI_{\tau-1}(0,0)}\biggl|_{T=T_c}
\nonumber \\%
&\sim&\left(\frac{\partial\Sigma^-}{\partial\mu}-\frac{\partial\Sigma^+}{\partial\mu}\right)_{T=T_c} \, ,
\end{eqnarray}
for  $\tau>2$ since in this case the integrals $I_{\tau-\varkappa}(0,0)$ and $I_{\tau-1}(0,0)$ remain finite.
 Therefore, the temperature behavior of  $\Delta\rho$ is defined by the difference of the reduced surface tension coefficient partial derivatives calculated  below and above the PT line. 
 We would like to pay an attention to an  analysis of the condition $\Delta\rho\ge0$. 
Clearly,  this condition simply means that the baryonic density increases during the deconfinement PT  
from hadrons to QGP, since the baryonic density of  latter is higher.

 Since the integral $I_{\tau-\varkappa}(0,0)$ and  the denominator staying in the expression after the second equality sign in  (\ref{EqXXVII})  is positive, then the condition $\Delta\rho\ge0$  is provided by the inequality  
$\frac{\partial\Sigma^+}{\partial\mu}<\frac{\partial\Sigma^-}{\partial\mu}$, which is the case for $\zeta^+=1$ only. Therefore, according to  the  parameterizations (\ref{EqIX})  and  (\ref{EqXIII}) one obtains 
%
%%%%%%%%%%%%%%%%%%%%%   Beta   %%%%%%%%%%%%%%%%%%%%%%%%%%%%%%%%%%%%%%%%%%%%%%%%%%%%%%%%%%%%%%%%%%%
%
\begin{eqnarray}
\label{beta}
\beta=
\left\{\begin{array}{lr}
\beta^+,  &\hspace*{0.1cm}
{\rm for} \quad \zeta^->\zeta^+=1 \, ,\\
{\rm min}(\beta^+,\beta^-),  &\hspace*{0.1cm}
{\rm for} \quad \zeta^-=\zeta^+=1  \, ,
\end{array}\right. 
\end{eqnarray}
where the notation $\beta^\pm \equiv \zeta^\pm + \xi^\pm - \frac{1}{\xi^T} $ is used. 
Since  in accord with  the adopted assumptions  the densities of pure phases are finite at the CEP, it follows that the index $\beta$ is non-negative.

%%%%%%%%%%%%%%%%%%%%%   Gamma'   %%%%%%%%%%%%%%%%%%%%%%%%%%%%%%%%%%%%%%%%%%%%%%%%%%%%%%%%%%%%%%%%%

To find  the critical exponent $\gamma'$ one has to calculate the  isothermal compressibility $K_T\equiv\frac{1}{\rho}(\frac{\partial\rho}{\partial p})_T$ for both pure  phases. Using  the baryonic density definition one can rewrite the isothermal compressibility as $K_T=\frac{1}{\rho^2}\frac{\partial^2p}{\partial\mu^2}$. Therefore, using Eq. (\ref{EqX}) at  the CEP  and keeping   the  most
singular terms one finds
\begin{eqnarray}
\label{EqXXIX}
\Delta K_T\hspace*{-0.2cm}&\simeq&\hspace*{-0.2cm}-\left(
\frac{\frac{\partial^2\Sigma^-}{\partial\mu^2}}{{\rho^-}^2}-\frac{\frac{\partial^2\Sigma^+}{\partial\mu^2}}{{\rho^+}^2}\right)\hspace*{-0.1cm}\cdot\hspace*{-0.1cm}
\frac{TuI_{\tau-\varkappa}(0,0)}{1-\frac{\partial F_H}{\partial z}+uI_{\tau-1}(0,0)}\biggl|_{T=T_c}
\nonumber \\
&\sim&\hspace*{-0.2cm}-\left(\frac{\frac{\partial^2\Sigma^+}{\partial\mu^2}}{{\rho^+}^2}-
\frac{\frac{\partial^2\Sigma^-}{\partial\mu^2}}{{\rho^-}^2}\right)_{T=T_c}.
\end{eqnarray}
Such a quantity, as suggested  in \cite{Fi:70},  is rather convenient for the index $\gamma'$ evaluation. 
The  detailed derivation of the expression  (\ref{EqXXIX}) can be found  in the Appendix A. 
 Indeed, differentiating Eq. (\ref{EqXXVII}) with the help of the parameterization  (\ref{EqXIII})   and recalling that its left hand side is just $|t|^\beta$, one immediately  finds 
\begin{equation}\label{gamma}
\gamma'=\frac{1}{\xi^T}-\beta.
\end{equation}
 It is evident, that the direct way to calculate this index via the isothermal compressibilities 
$K_T$ of pure phases  would give us the same result for the present model.
Thus, 
at the  first glance it seems that the definition  (\ref{EqXX}) suggested in \cite{Fi:70} is rather appropriate for the case 
of the asymmetric phase diagram, when the $\gamma'$  indices of two pure phases are different. We, however, will  return to  this question again  in the subsequent section.

%%%%%%%%%%%%%%%%%%%%%   Delta   %%%%%%%%%%%%%%%%%%%%%%%%%%%%%%%%%%%%%%%%%%%%%%%%%%%%%%%%%%%%%%%%%%

To calculate  the critical exponent $\delta$ one can use  the definition of $\Delta z$
at  the critical isotherm  and  get
\begin{eqnarray}\label{EqXXXI}
\widetilde{p}-p_{cep}&=&T_{cep}(\widetilde{\Delta} z+\widetilde{z}_M-z_M |_{cep})
\nonumber \\
&=&T_{cep}\left(\widetilde{\Delta} z+\Delta\mu\frac{\partial z_M}{\partial\mu}\biggl|_{cep}\right),
\end{eqnarray}
where in the second step one has to expand  $\widetilde{z}_M$  in powers of $\Delta\mu \equiv \mu - \mu_{cep}$ and keep the linear  term. Similarly one can  determine  the deviation of the   baryonic density  at the critical isotherm from that one at the CEP. 
Note that this  critical isotherm  is  lying  outside the mixed phase  and necessarily it belongs to the high density phase. Then one finds 
\begin{eqnarray}\label{EqXXXII}
\widetilde{\rho}-\rho_{cep}&=&T_{cep}\left(\frac{\partial\widetilde{\Delta} z}{\partial\mu}+
\frac{\partial\widetilde{z}_M}{\partial\mu}-\frac{\partial z_M}{\partial\mu}\biggl|_{cep}\right)
\nonumber \\
&=&T_{cep}\left(\frac{\partial\widetilde{\Delta} z}{\partial\mu}+
\Delta\mu\frac{\partial^2z_M}{\partial\mu^2}\biggl|_{cep}\right).
\end{eqnarray}
Analysis of the functions  $\widetilde{\Delta} z$ and $\frac{\partial\widetilde{\Delta} z}{\partial\mu}$ near the CEP shows that  it is convenient to substitute the expansion
$I_\tau(\widetilde{\Delta} z,\widetilde{\Sigma}) \approx I_\tau(0,0)-
\widetilde{\Delta} zI_{\tau-1}(\frac{\widetilde{\Delta} z}{2},\frac{\widetilde{\Sigma}}{2})-
\widetilde{\Sigma}I_{\tau-\varkappa}(\frac{\widetilde{\Delta} z}{2},\frac{\widetilde{\Sigma}}{2})$
\cite{Bugaev_07_physrev} into Eq. (\ref{EqX})  for the  cross-over states  in order to
obtain
\begin{equation}\label{EqXXXIII}
\widetilde{\Delta} z=\frac
{\Delta\mu A_\mu |_{cep}
-~\widetilde{\Sigma}\, \widetilde{u}\, I_{\tau-\varkappa}(\frac{\widetilde{\Delta} z}{2},\frac{\widetilde{\Sigma}}{2})}
{1-\frac{\partial\widetilde{F}_H}{\partial\mu}+
\widetilde{u}\,I_{\tau-1}(\frac{\widetilde{\Delta} z}{2},\frac{\widetilde{\Sigma}}{2})} \,,
\end{equation}
where it is sufficient to  keep only the first order terms of the expansion and to  use
 the fact that at the CEP $\widetilde{\Delta} z=0$. 
According to Eq. (\ref{EqXXXI}) the higher order terms of expansion could be neglected. 

Using the same steps of derivation  one can show that
\begin{equation}\label{EqXXXIV}
\frac{\partial\widetilde{\Delta} z}{\partial\mu}=\frac
{\Delta\mu \frac{\partial A_\mu}{\partial\mu} |_{cep}
-~\frac{\partial\widetilde{\Sigma}}{\partial\mu}\, \widetilde{u}\, I_{\tau-\varkappa}(\frac{\widetilde{\Delta} z}{2},\frac{\widetilde{\Sigma}}{2})}
{1-\frac{\partial\widetilde{F}_H}{\partial\mu}+
\widetilde{u}\,I_{\tau-1}(\frac{\widetilde{\Delta} z}{2},\frac{\widetilde{\Sigma}}{2})} \,.
\end{equation}
The reduced surface tension coefficient partial derivative $\frac{\partial\widetilde{\Sigma}}{\partial\mu}$ vanishes at the CEP, whereas its second partial derivative $\frac{\partial^2\widetilde{\Sigma}}{\partial\mu^2}$ diverges at this point. Therefore,  from Eqs. (\ref{EqXXXI}), (\ref{EqXXXIII}), (\ref{EqXXXII}) and  (\ref{EqXXXIV}) one finds that
$\widetilde{\Delta} z\sim\Delta\mu$ and
$\frac{\partial\widetilde{\Delta} z}{\partial\mu}\sim\frac{\partial\widetilde{\Sigma}}{\partial\mu}$. Then 
using   Eqs. (\ref{EqIX}) and  (\ref{EqXIII}) one concludes that
\begin{equation}\label{delta}
\delta=\frac{1}{\xi^T\beta^+}.
\end{equation}
Note that the  condition $\delta>1$ requires $\frac{1}{\xi^T} > \beta^+ $ which according to the result found  for the index $\gamma'$ (\ref{gamma}) is  consistent with the  constrain $\gamma'>0$.

%%%%%%%%%%%%%%%%%%%%%%%%%%%%%%%%%%%%%%%%%%%%%%%%%%%%%%%%%%%%%%%%%%%%%%%%%%%%%%%%%%%%%%%%%%%%%%%%%%%%%%%%%%%%%%%%%%%%%%%%

Despite some  similarities the  critical exponents of the QGBSTM2 are  very different from  the critical indices  of  the comparable cluster   models such as the FDM \cite{Fisher_67}, the SMM \cite{Reuter_01} and the QGBSTM1 \cite{Ivanytskyi}. 
Thus, 
surprisingly, the critical exponents of the present model do not depend on the Fisher topological exponent $\tau$ and  on  the  parameter $\varkappa$ which  relates    the mean surface of the QGP bag to its volume. 
It seems to be a unique feature of the QGBSTM2, since the critical exponents of the FDM, SMM and QGBSTM1 are expressed in terms  these parameters. 
Furthermore, in contrast to the QGBSTM1 critical exponents \cite{Ivanytskyi},   the  parameter $\chi$  does not appear in any final  expression for the critical indices of the QGBSTM2. This is another  surprising fact  since just this  parameter switches  the  different regimes  between  the QGBSTM1 critical exponents \cite{Ivanytskyi}.

Having the explicit  expressions for the QGBSTM2 critical exponents 
 let us now  discuss the question  whether the obtained results  are able to reproduce the indices of the 2-dimensional Ising model \cite{Huang}, of   the 3-dimensional Ising model \cite{Campostrini:05}, of the simple liquids \cite{Huang} and 
 of the  O(4) symmetric 3-dimensional spin model \cite{CritExponO4,CritExponO4B} (see the Table \ref{table1}). It is an important  question since, as it is expected, the universality class of the 3-D Ising model coincides  with that one of the 3-flavor degenerated QCD, whereas 
 the 2+1 flavor QCD  falls into  universality class of
 the O(4) spin  model
 \cite{Rob:84, Wilczek:92, Karsch:01, Misha:07}.

As one can see  from the Table \ref{table1}  $\gamma'>1$ for all discussed systems. 
On the other hand, from the explicit expressions of the QGBSTM2  critical exponents one can see that 
within the present model  $\gamma'>1$ for $\xi^T<1$ only,  which immediately  leads to a conclusion that $\alpha'=0$.
Therefore, the  QGBSTM2 with the traditional definition of index $\alpha'$ is able to reproduce the critical exponents of the 2-dimensional Ising model only. 
Such a problem is not a  new one. For example, the critical exponents of the SMM \cite{Reuter_01} do not reproduce that ones of the simple liquids and of the  3-dimensional Ising model since $\alpha'_{SMM}=0$. It is, however,  believed that such a problem is related to the traditional  definition of the critical exponent $\alpha'$. In order to elucidate this fact let us study in detail  the scaling relations for the indices of  the present model. 

\begin{table}[t]
\bigskip
\begin{tabular}{|c|c|c|c|c|c|c|c|c|c|c|c|c|c|}
\hline
     \hspace*{1.0cm} &     2d Ising         &       Simple      &       3d Ising         &   O(4) spin   \\ 
                                  &        model         &       liquids     &         model           &     model      \\ \hline
            $\alpha'$       &             0            &      0.10(1)     &      0.1096(5)      &    -0.19(6)   \\  \hline
             $\beta$        &   $\frac{1}{8}$  &    0.335(15)   &      0.3265(1)      &     0.38(1)   \\  \hline
           $\gamma'$     &   $\frac{7}{4}$  &     1.25(5)      &      1.2373(2)      &     1.44(4)   \\  \hline
              $\delta$      &            15           &       4.5(3)      &      4.7893(8)      &     4.82(5)   \\  \hline
\end{tabular}
\vspace*{0.3cm}
\caption{
The critical indices of the  2-dimensional Ising model \cite{Huang}, of the simple liquids \cite{Huang},  of the 3-dimensional Ising model \cite{Campostrini:05} and of the O(4) symmetric spin model \cite{CritExponO4,CritExponO4B}.
}
\label{table1}
\end{table}

%%%%%%%%%%%%%%%%%%%%%   Scaling relations    %%%%%%%%%%%%%%%%%%%%%%%%%%%%%%%%%%%%%%%%%%%%%%%%%%%%%

%
\section{The Scaling relations of the QGBSTM2}
\label{secscallings}

Now we return to the discussion of the scaling  inequalities (\ref{F1}), (\ref{G}) and (\ref{L}) of the present model. This inequalities are well established analytically \cite{Fi:64, Gri:65, Li:66} and experimentally \cite{Stanley:71,Huang,Lavis:99}. As one can see from the Table II 
these inequalities are exactly established only  for an exact solution of the 2-dimensional Ising model while for other systems shown there   such inequalities  are established within the  error bars, which in some cases are not very small. 
From time to time in the literature there appear the models  \cite{Fi:70, Reuter_01, Ivanytskyi} and  even the 
experimental works  (we mean the text-book example  \cite{Roach:68} discussed in the quoted  edition of \cite{Stanley:71}) in which the  problems related to the Fisher 
(\ref{F1}) and the Griffiths (\ref{G}) inequalities  are reported. It is remarkable that the reported problems are always related to the inequalities in which  the exponent $\alpha'$ is involved.  The situation is somewhat mysterious since the formal conditions of the well known Fisher theorem \cite{Fi:64} proving the validity of these inequalities for liquids are fulfilled by  the models \cite{Fi:70, Reuter_01, Ivanytskyi}, but there is the range of parameters for which either one of the relations (\ref{F1}) and (\ref{G}) or both of them are not obeyed. Therefore, it is interesting to verify the scaling relations for the indices of the QGBSTM2 which, as we discussed in the preceding sections, demonstrate the markable difference with that ones of the models  \cite{Fi:70, Reuter_01, Ivanytskyi}.

The explicit expressions for the critical exponents found above allow us to directly examine the scaling inequalities. Despite the usual  expectations, the  Fisher and Griffiths  inequalities are not always  obeyed, whereas the Liberman  inequality is fulfilled for any values of the model parameters. 
Let's   first demonstrate the validity of the Liberman inequality (\ref{L}). Using the explicit expressions for the indices  $\beta$,  $\gamma'$ and $\delta$, i.e. Eqs. (\ref{beta}), (\ref{gamma}) and (\ref{delta}),  one obtains
\begin{eqnarray}\label{EqXXXVI}
%
%\hspace*{-.25cm}
&\gamma'+\beta(1-\delta)=\frac{1}{\xi^T}\left(1-\frac{\beta}{\beta^+}\right)
\nonumber \\
&=
\left\{\begin{array}{lr}
0,  &%\hspace*{0.1cm}
{\rm for}\hspace*{-0.2cm} \quad \zeta^->\zeta^+=1\\
\frac{1}{\xi^T}\left(1-\frac{{\rm min}(\beta^+,\beta^-)}{\beta^+}\right)\ge0,  &%\hspace*{0.1cm}
{\rm for} \hspace*{-0.2cm}\quad \zeta^-=\zeta^+=1
\end{array}\right.,\hspace*{0.2cm}
\end{eqnarray}
where an obvious inequality ${\rm min}(\beta^+,\beta^-)\le\beta^+$ is accounted for. 
From the  Liberman  scaling law  (\ref{L}) one immediately deduces  the relation between the scaling  inequalities (\ref{F1}) and (\ref{G}) $\alpha'+\beta(\delta+1)\le\alpha'+2\beta+\gamma'$. Then using the explicit expressions for  the QGBSTM2  critical exponents  one  can get the following result for the Fisher and Griffiths inequalities
\begin{eqnarray}\label{EqXXXVII}
\hspace*{-.1cm}
&\alpha'+\beta(\delta+1)\le\alpha'+2\beta+\gamma'
\nonumber \\
&=-2\left(\min(1,\frac{1}{\xi^T})-1-\frac{1}{\xi^T}\right)-\left(\frac{1}{\xi^T}-\beta\right)
\nonumber \\
&
=2\max(1,\frac{1}{\xi^T})-\gamma'\,.
\end{eqnarray}
Note that in the evaluation of the second equality above   we employed the expression for index $\gamma'$ and the following sequence of steps  $\min(1,\frac{1}{\xi^T})-1-\frac{1}{\xi^T}=\min(-1,-\frac{1}{\xi^T})=-\max(1,\frac{1}{\xi^T})$.  Eq. (\ref{EqXXXVII}) gives the estimate from above for the left hand side of the Griffiths inequality (\ref{G}) and also it demonstrates that the scaling laws 
(\ref{F1}) and  (\ref{G}) may  be broken for the QGBSTM2, if  $2\max(1,\frac{1}{\xi^T})-\gamma' < 2$.   
Moreover, Eq. (\ref{EqXXXVII})  clearly shows that, if   the Fisher scaling inequality is fulfilled,  this   does not guaranty that the Griffiths one is obeyed. 
One of the consequences of Eq. (\ref{EqXXXVII}) is an appearance of the following  relation between the QGBSTM2 critical exponents  
\begin{equation}\label{EqXXXVIII}
\alpha'+2\beta+2\gamma'\ge2\,,
\end{equation}
which is based on an obvious inequality $2\max(1,...)\ge2$. Note also  that for $\xi^T\ge1$ the  inequality  (\ref{EqXXXVIII}) turns into an  equality since $\max(1,\frac{1}{\xi^T})=1$ in this case.

\begin{table}[t]
\bigskip
%
%\hspace*{-.5cm}
%
%
%
\begin{tabular}{|c|c|c|c|c|c|c|c|c|c|c|}
\hline
 \begin{tabular}{c}
                          \\
                          \\ \hline
 $\alpha'+2\beta+\gamma'$ \\ \hline
 $\alpha'+\beta(\delta+1)$\\ \hline
 $\gamma'+\beta(1-\delta)$
\end{tabular}                       &\begin{tabular}{c}
                                    Ising model    \\ \hline
                                      \begin{tabular}{c|c}
                                      2D &       3D       \\ \hline
                                      2  &   1.99996(7)   \\ \hline
                                      2  & 2.000412(5000) \\ \hline
                                      0  & -0.000052(2000)
                                      \end{tabular}
                                     \end{tabular}                  & \begin{tabular}{c}
                                                                        Simple  \\
                                                                        liquids  \\ \hline
                                                                       2.0200(55)   \\ \hline
                                                                       1.9425(55)   \\ \hline
                                                                       0.0775(212)
                                                                    \end{tabular}
                                                                                             & \begin{tabular}{c}
                                                                                                      O(4)     \\
                                                                                                      model      \\ \hline
                                                                                                     2.01(8)  \\ \hline
                                                                                                     2.02(9)  \\ \hline
                                                                                                    -0.01(6)
                                                                                             \end{tabular}
\\ \hline
\end{tabular}
\vspace*{0.3cm}
\caption{
Scaling relations between the critical exponents taken  from  the Table \ref{table1}. The uncertainties  were calculated from their values given in the Table \ref{table1} using the error determination method for indirect measurements \cite{BookOnEr}.
}
\label{table2}
\end{table}

In order to recover  the scaling inequalities (\ref{F1}) and (\ref{G}) we 
follow  the  suggestion of  \cite{Fi:70} and  replace the index $\alpha'$ by $\alpha'_s$, where the latter   describes the temperature dependence of the specific heat capacity difference of two phases at the 
CEP, i.e. 
$\Delta C=(C_{\rho^+}-C_{\rho^-})_{T=T_c}$. 
The idea behind such a suggestion is to get the most singular term from the difference of the specific heats capacity of two pure phases.  
Then from the  Clapeyron-Clausius equation one can find that
\begin{eqnarray}\label{EqXXXIX}
\Delta C&=&T_c\frac{\rho^--\rho^+}{\rho^+\rho^-}\frac{d}{dT}\left(s_M+\frac{d\mu_c}{dT}\rho_M\right)
\nonumber \\
&-&T_c\frac{d\mu_c}{dT}\frac{d}{dT}\ln\frac{\rho^+}{\rho^-} \,.
\end{eqnarray}
Since $\rho^+-\rho^-\sim t^\beta$ in the vicinity of the CEP, then using   (\ref{EqXIII}) and the fact that the function $z_M (T, \mu)$ together  with all  its derivatives up to the second one are finite,  we  conclude that the first term on the right hand side of  (\ref{EqXXXIX}) behaves as $t^{\min(0,\frac{1}{\xi^T}-2)+\beta}$. 
Similarly one  finds  that the second term on the right hand side of  (\ref{EqXXXIX})  behaves as $t^{\frac{1}{\xi^T}-2+\beta}$. This  analysis shows that the first of the two discussed terms   is the leading one and, hence, we have  
\begin{eqnarray}\label{EqXXXX}
\alpha'_s=
\max\left(2,\frac{1}{\xi^T}\right)-\beta-\frac{1}{\xi^T}.
\end{eqnarray}
Note  that $\alpha'_s\ge0$  for $\frac{1}{\xi^T}<2-\beta$ only. Using  the  $\alpha_s'$ expression  (\ref{EqXXXX})  one can get 
\begin{equation}\label{EqXXXXI}
\alpha'_s+2\beta+\gamma'=\max\left(2,\frac{1}{\xi^T}\right)\ge2,
\end{equation}
which is in a complete  agreement with the expectation of  \cite{Fi:70}. However, such a replacement  does not recover the
Griffiths inequality. Indeed, in this case one finds
\begin{equation}\label{EqXXXXII}
\alpha'_s+\beta(1+\delta)=2-\min\left(2,\frac{1}{\xi^T}\right)+\frac{\beta}{\xi^T\beta^+} \,. 
\end{equation}
Obviously, the right hand side of (\ref{EqXXXXII}) is less than $2$ for $\beta=\beta^-<\beta^+\min(1,2\xi^T)$.
Thus, the QGBSTM2 gives the first and  explicit example that the definition of the $\alpha'_s$
index does not recover the Griffiths scaling inequality, although it, indeed, redeems the Fisher 
scaling law. The reason that  the $\alpha'_s$ definition worked well in all preceding models, but fails for this one, is that  within the  models \cite{Fi:70, Reuter_01, Ivanytskyi} the  PT corresponds to 
the change of the leading singularity from the simple pole    of the isobaric partition partition to its 
essential singularity, whereas in the QGBSTM2 the leading singularity above and below PT is the simple pole.  Then for the models  \cite{Fi:70, Reuter_01, Ivanytskyi} the divergent terms in the difference $\Delta C=(C_{\rho^+}-C_{\rho^-})_{T=T_c}$ did not cancel each other because they have 
rather different analytical  behavior, while in  the QGBSTM2  the most divergent terms in the specific heat capacity  of two phases  coincide with each other  and they simply  cancel each other in the expression for $\Delta C$. Therefore,  the  QGBSTM2 provides   us with  a direct evidence that  such a simple and attractive definition of the $\alpha'_s$ index \cite{Fi:70}  which, so far, was designed   for the asymmetric phase diagrams and worked well for  the models  \cite{Fi:70, Reuter_01, Ivanytskyi}, should be 
replaced now by a proper one.

Keeping in mind these facts we introduce a new definition for the index $\alpha'$ 
which is based on the behavior of the  linear combination of specific heat capacity of two phases $C_{tot} \equiv \left[K^+C^++K^-C^-\right]_{T=T_c}\sim |t|^{-\alpha'_c}$ with the positive and nonvanishing  coefficients  $K^\pm > 0$ which in general may  depend  on $T$ or $\mu$.
The inequality  $\frac{K^+}{K^-}>0$ guarantees that no  term in  $C_{tot}$ is missing or cancelled. 
On the other hand
the condition $K^\pm|_{T=T_c} > 0$ provides that the index $\alpha'_c$  has the maximal value among that one of pure phases.  Note that this condition leads to a difference with  the traditional 
definition of  the index $\alpha'$ since the coefficients 
$K^+|_{T=T_c}$  and $K^-|_{T=T_c}$ are not related to the fractions of  pure  phases. Such a property allows us to avoid the 
situation, when the  curve at which the specific heat capacity is found does not match the boundary with one of the phases which was the case for the models analyzed in \cite{Fi:70, Reuter_01}.

The  index $\alpha'_c$ definition has a clear  physical interpretation:
when approaching the CEP the density fluctuations  get so strong that the specific heat capacity of  both phases contribute into the measurable value and, hence, the largest term determines the critical exponent $\alpha'_c$. Then the coefficients $K^+|_{T=T_c}$  and $K^-|_{T=T_c}$  define the  weight of corresponding  pure phase into the specific heat capacity of their mixture. 
Obviously, the maximization of the  $\alpha'$-value is not well suited for the pure phases with asymmetric   
properties, but also it should  increase the magnitude of the  left hand side of  the Fisher and Griffiths inequalities and, thus, it should weaken the necessary conditions  to obey them.

Let's find out the index $\alpha'_c$.  Making  the same steps as  in deriving Eq. (\ref{EqXXXIX}) we obtain
\begin{eqnarray}
C_{tot} &= & %\left[K^+C^++K^-C^-\right]_{T=T_c}=
T\left(\frac{K^+}{\rho^+}+\frac{K^-}{\rho^-}\right)\frac{ds_M}{dT}\bigl|_{T=T_c}
\nonumber
\\
\label{EqXXXXIII}
&-& T\left[\frac{K^+}{\rho^+}\frac{d}{dT}\left((\rho^+-\rho_M)\frac{d\mu_c}{dT}\right)\right.
\nonumber \\
&+&\left.\frac{K^-}{\rho^-}\frac{d}{dT}\left((\rho^--\rho_M))\frac{d\mu_c}{dT}\right)\right]_{T=T_c}. 
\end{eqnarray}
From this formula one can draw  an  important conclusion that negative values of the index  $\alpha'_c$ are possible, if  and only if 
each full $T$-derivative on the right hand side of (\ref{EqXXXXIII})
including $\frac{d\mu_c}{dT}\bigl|_{cep}$ and  $\frac{ds_M}{dT}\bigl|_{T=T_c}$  vanish at  the CEP.
The necessary conditions for $\alpha'_c < 0$ are $\frac{d\mu_c}{dT}\bigl|_{cep} = 0$ and 
\begin{equation} \label{EqXXXXIV}
\frac{\partial s_M}{\partial T}\bigl|_{T=T_c}\sim |t|^\omega ,
\end{equation}
where $\omega>0$ for $\alpha'_c<0$ and $\omega=0$ for $\alpha'_c\ge0$. The  geometrical meaning of  Eq. (\ref{EqXXXXIV}) is that the function $s_M(T,\mu_{cep})$ has a kink point at the critical temperature for $\omega>0$.  The case $\omega=0$ does not add anything new to the  above  evaluation of  the indices $\alpha'$ and $\alpha'_s$.
Using the parameter $\omega$ one 
can show that the first term on the right hand side of  Eq. (\ref{EqXXXX}) behaves as $t^{\min(\omega,\frac{1}{\xi^T}-1)}$. Since the coefficients $K^\pm|_{T=T_c}$ are positive, then each term  $\frac{K^\pm}{\rho^\pm}\frac{d}{dT}\left((\rho^\pm-\rho_M)\frac{d\mu_c}{dT}\right)$ in (\ref{EqXXXXIV}) should vanish independently. 
As shown in the Appendix B, 
employing the parameterizations (\ref{EqIX}) and (\ref{EqXVII}) 
one can find  that  $\max[ (\rho^+-\rho_M), ~ (\rho^- -\rho_M)] \sim t^\beta$ at the vicinity of the CEP.  
Taking into account this  result  one can deduce  that  the term staying inside   the square brackets on the right hand side of   Eq. (\ref{EqXXXXIII}) behaves as $t^{\frac{1}{\xi^T}+\beta-2}$. Thus, this analysis allows us to determine the critical exponent $\alpha'_c$ as 
\begin{eqnarray}\label{EqXXXXV}
\alpha'_c=
\max\left(\frac{1}{\xi^T}+\beta-\omega,2\right)-\beta-\frac{1}{\xi^T}.
\end{eqnarray}

Similarly   the exponents $\alpha'$ and $\alpha'_s$   can be  found for the nonzero  values of $\omega$ 
\begin{eqnarray}\label{EqXXXXVI}
\alpha'|_{\omega\neq0} & = & 2-2\min\left(\frac{1}{\xi^T},1+\frac{\omega}{2},\frac{1+\frac{1}{\xi^T}}{2}\right) \,,
\\
\label{EqXXXXVII}
\alpha'_s|_{\omega\neq0} & = & \max\left(\frac{1}{\xi^T}-\omega,2\right)-\beta-\frac{1}{\xi^T} \, .
\end{eqnarray}
Comparing Eqs. (\ref{EqXXXXV}) and (\ref{EqXXXXVII}) one can easily see that $\alpha'_c\ge\alpha'_s$ for the same value of  $\omega$. Moreover, for the case $\frac{1}{\xi^T}-\omega > 2$ one finds that  $\alpha'_c = \alpha'_s + \beta = - \omega$, i.e. the index $\alpha'_c$ can be essentially larger then $\alpha'_s$  since $\beta > 0$. 
Therefore, it is evident that in the models \cite{Fi:70, Reuter_01, Ivanytskyi} in which 
the $\alpha'_s$ index recovers the scaling inequalities (\ref{F1})--(\ref{L}) the $\alpha'_c$ index should  recover  them too, since by the construction it is not smaller than $\alpha'_s$, but 
for other models the properties of this index should be studied.

Let us demonstrate  explicitly that  the proposed definition of the critical exponent 
$\alpha'_c$  is  more adequate with respect to the QGBSTM2 scaling inequalities.
From the expression (\ref{gamma}) for the index $\gamma'$  one gets  $\frac{1}{\xi^T}=\gamma'+\beta$. Substituting this result into (\ref{EqXXXXV}) one finds 
\begin{eqnarray}\label{EqXXXXVIII}
\hspace*{-0.3cm}\alpha'_c  & = & 
\left\{
\begin{tabular}{ll}
$-\omega$ \,, \hspace*{1.0cm} {\rm for} ~ $\omega~\le~\gamma'+2\beta-2$~, \\
\\
$2-2\beta-\gamma'$\,,\hspace*{0.0cm} {\rm for} ~ $\omega\ge\gamma'+2\beta-2=-\alpha'_c$~,
\end{tabular}
\right.
\end{eqnarray}
%
%%$\alpha'_c=-\omega$ for $\omega\le\gamma'+2\beta-2$ or $\alpha'_c=2-2\beta-\gamma'$ for $\omega\ge\gamma'+2\beta-2=-\alpha'_c$. 
%
and, hence, $\alpha'_c \ge 2-2\beta-\gamma'$ which is just  the Fisher inequality written for  the  $\alpha'_c$ exponent.  
Note that in contrast to the $\alpha'_c$ definition  the Fisher inequality for  the $\alpha'_s$ definition   is not fulfilled anymore.  For instance, 
if $\alpha'_c = - \omega$, then from Eqs.  (\ref{EqXXXXVII}) and  (\ref{EqXXXXVIII}) it follows that  
for the same value of  the index $\omega$ the relation $\alpha'_s + 3\beta + \gamma' \ge 2$ holds  which, obviously,  differs  from the Fisher inequality. The considered example clearly demonstrates that for $\alpha' <0$ the most general definition of the index $\alpha'_s$ leads to the difficulties with the Fisher scaling relation and, hence, it has to be replaced by the $\alpha'_c$ definition. 

Since the Liberman inequality  (\ref{EqXXXVI}) does not depend on the $\alpha'$ definition, it holds, but here 
we would like to pay an attention to the case when it turns  into an  equality. From Eq. (\ref{EqXXXVI})  one concludes that 
this occurs for $\beta=\beta^+$, which applied to the relations for the indices  $\delta$ and $\gamma$ yields 
 a very important relation
\begin{eqnarray}\label{EqXXXXIX}
  \gamma'+\beta=\beta\, \delta = \frac{1}{\xi^T} \,,
\end{eqnarray}
that should hold  for any model  in which the scaling relation (\ref{L}) is obeyed as equality, for instance, for $O(2)$, $O(3)$ and $O(4)$ spin models.  For such models the exponent $\xi^T = \frac{y_h}{y_t}$ is the ratio of  the so-called  magnetic $y_h$ and thermal $y_t$ exponents \cite{Lavis:99, Karsch:10, Karsch:11} and it defines the curvature radius  of the phase diagram  in  $\mu-T$ plane \cite{ Karsch:10}.  If, however,   the scaling law (\ref{L}) is obeyed as an inequality, the relation (\ref{EqXXXXIX}) reads as an  inequality too, i.e. $  \beta\, \delta  \, \xi^T <  1$. 

For the Griffiths inequality (\ref{G}) the direct substitution of the expressions (\ref{beta}), (\ref{gamma}), (\ref{delta}) and 
(\ref{EqXXXXV}) yields 
\begin{eqnarray}\label{EqXXXXX}
%\hspace*{-0.2cm}
&&\alpha'_c+\beta(1+\delta) =\nonumber \\
&&=
\left\{\hspace*{-0.1cm}\begin{array}{lr}
-\omega+\beta(1+\delta),  &\hspace*{-0.1cm}
{\rm for}~ \omega\le\beta+\frac{1}{\xi^T}-2 \,,\\
2-\frac{1}{\xi^T}\left(1-\frac{\beta}{\beta^+}\right),  &\hspace*{-0.1cm}
{\rm for}~ \omega\ge\beta+\frac{1}{\xi^T}-2\, .
\end{array}\right.
\end{eqnarray}
Note that  the right hand side of Eq. (\ref{EqXXXXX}) can be smaller than 2 in two cases. 
Indeed,   if the index $\omega$  obeys the inequalities $\beta(1+\delta)-2<\omega<\beta+\frac{1}{\xi^T}-2$, then   the 
Griffiths scaling law (\ref{G})    is broken down. The other possibility for the break down of this 
scaling law corresponds to  the lower equality in (\ref{EqXXXXX}), i.e. for 
\begin{equation}\label{EqXXXXXIa}
\alpha'_c+\ \beta (1 +  \delta) =  2 - \delta (\beta^+  - \beta) \, ,
\end{equation}
that  occurs  for $\omega\ge\beta+\frac{1}{\xi^T}-2$ and $\beta < \beta^+$. Thus, from these examples  one might conclude that even the improved definition (\ref{EqXXXXV}) of the index $\alpha'$ which perfectly works for the Fisher inequality  does not save the Griffiths scaling law. 
We, however, would like to stress that  these examples clearly show us that the  problem of formulating the scaling inequalities for the asymmetric phase diagrams is much deeper and cannot 
be resolved by the redefinition of the indices. It also requires the modification of traditional scaling relations. 
Indeed, using the Fisher inequality with the  $\alpha'_c$ exponent  and the explicit expressions  for the indices $\gamma$ and $\delta$ we obtain another inequality 
\begin{equation}\label{EqXXXXXII}
\alpha'_c+\beta + \beta^+ \delta \ge 2 \, ,
\end{equation}
which is an analog of the Griffiths inequality for the QGBSTM2  exponents. 

Let us show that the scaling law  (\ref{EqXXXXXII}) cannot be broken within the QGBSTM2.
For this purpose  we assume  that $\alpha'_c+\beta + \beta^+ \delta \equiv   \alpha'_c+\beta + \frac{1}{\xi^T}  < 2$. Then, comparing the latter  assumption with  the upper equality in (\ref{EqXXXXX}), we arrive at the contradiction, since in this case $\omega = - \alpha'_c \le  \beta + \frac{1}{\xi^T} -2 $, but our assumption is equivalent to the different  inequality, i.e.  $\omega = - \alpha'_c >   \beta + \frac{1}{\xi^T} -2 $.  Analogously, from the lower expression in  (\ref{EqXXXXX}) it follows that $\alpha'_c +\beta  +
\beta^+ \delta = 2$ and we again obtain the contradiction with the original assumption which, as we proved,  is the false one. 

From  Eq. (\ref{EqXXXXX}) one can deduce that Eqs.   (\ref{EqXXXXIX}) and  (\ref{EqXXXXXIa}) hold as equalities   for  
$\omega\ge\beta+\frac{1}{\xi^T}-2$ and 
   $\beta = \beta^+$,      and, hence,  
the Fisher scaling relation (\ref{F1}) becomes an equality too. 
Thus, in this case all scaling laws (\ref{F1})--(\ref{L}) are fulfilled as equalities and this is the traditional scaling  regime. 
In all other cases defined by Eq. (\ref{EqXXXXX}), i.e. for  $\beta(1+\delta)-2<\omega<\beta+\frac{1}{\xi^T}-2$ or for $\omega\ge\beta+\frac{1}{\xi^T}-2$ and $\beta < \beta^+$, 
the Fisher and Liberman scaling laws are obeyed as inequalities, but the Griffiths one in its usual form  (\ref{G}) is broken down and, hence,  we suggest to use the inequality 
(\ref{EqXXXXXII}) instead of (\ref{G}). 
Indeed,  the latter, as we proved above,  is valid in more general case than the original  Griffiths one  and also
it seems to be more natural, since, in contrast to (\ref{G}), for  $\beta^+  > \beta = \beta^-$ the high density phase exponent $\delta$ in (\ref{EqXXXXXII})  is multiplied by other  high density phase  exponent  $\beta^+$.  Since in this regime the generalized inequality (\ref{EqXXXXXII}) is valid, we
call it the generalized scaling regime.

Also we have to stress that the  break down of the  traditional Griffiths inequality (\ref{G}) does not require some special conditions. Indeed, it is easy to see that, if the Fisher relation (\ref{F1}) is  fulfilled as equality and the Liberman one (\ref{L}) is  obeyed as a strict inequality, then the 
traditional Griffiths inequality (\ref{G}) would be broken, which is not the case for the inequality 
(\ref{EqXXXXXII}).  
Note that for the QGBSTM2  exponents  the Liberman inequality can be also generalized similarly to the Griffiths inequality (\ref{EqXXXXXII}). Indeed,  rewriting the first equality in   the expression  (\ref{EqXXXVI})  one finds
\begin{equation}\label{EqXXXXXIII}
\gamma' + \beta   - \beta^+  \delta =  0 \, ,
\end{equation}
which is stronger than the usual Liberman scaling inequality,  since it is possible that  $\beta^+ > \beta$. Here it is appropriate to discuss the definition (\ref{EqXX}) of the critical index $\gamma'$
suggested in \cite{Fi:70}. Similarly to the definition of the $\alpha'_s$ index \cite{Fi:70} such a definition is simple and  it may provide one with  the maximal value of the compressibility exponent of two pure phases.  However, it has the very same defect as the $\alpha'_s$ definition, namely it does not guaranty that for some models with the asymmetric phase diagrams  the leading terms of two compressibilities may simply  cancel each other 
as we found such a possibility   for the QGBSTM2 $\alpha'_s$ index and this would lead to the problems with the scaling relations.  Therefore, in order to avoid these problems  for such cases  we suggest to employ  the definition of the $\gamma'$ exponent which is similar to the $\alpha'_c$ definition (\ref{EqXXXXIII}) in which 
the heat capacities $C^\pm$ are replaced by the corresponding  thermal compressibilities $K_T^\pm$
\begin{equation}\label{EqXXXXXIIIb}
k^+ K_T^+ + k^- K_T^-  \sim |t|^{\gamma' } \, , 
\end{equation}
with  the positive and nonvanishing  coefficients  $k^\pm > 0$.  The interpretation of such coefficients is similar to that  one introduced for the $\alpha'_c$ index.

After the thorough discussion of the QGBSTM2 critical exponents 
it is not surprising that for the traditional scaling regime the present model  with the $\alpha'_c$ and $\omega$ definitions  is able to reproduce  the critical  indices of the 2-dimensional Ising model, of the simple liquids,  of the 3-dimensional Ising model  and of the O(4) symmetric spin model for which, as one can see from the Table II,  the inequalities (\ref{F1})--(\ref{L}) are well established. The list of the corresponding  QGBSTM2 parameters  is given in the Table III. 
However, 
an  existence of the generalized  scaling regime for  the QGBSTM2   with the most favorable definition of the  $\alpha'_c$ index for which the Griffiths  inequality is broken down  is rather surprising,  since  it is widely accepted that the validity of the Fisher (\ref{F1})  and the Griffiths  (\ref{G}) relations does not require any additional assumptions  except for the conditions of the Fisher theorem \cite{Fi:64,Stanley:71, Lavis:99}. Now   the QGBSTM2  provides us with an explicit example that this may not be the case for the Griffiths  inequality (\ref{G}) and the latter  has to be replaced by  (\ref{EqXXXXXII}).

It is worth to note that from the physical point of view the successful description of the critical exponents belonging to the
different universality classes by a single model not only shows that the QGBSTM2  is very general, but it also evidences for the fact  that 
the physical clusters or bags employed in this model are, indeed, the relevant degrees of freedom at the CEP for all analyzed systems, despite the  very different physics exhibiting by  the  original  Hamiltonians of  the corresponding  spin models. 
On the one hand this conclusion completes the finding  \cite{Moretto:05} that the FDM \cite{Fisher_67} correctly 
describes the distribution of large clusters of the 2- and 3- dimensional Ising model in the wide range of temperatures, including the CEP. On the other hand the  QGBSTM2  generalizes the results of the noninteracting cluster models \cite{Cluster:1,Cluster:2}  which quite successfully describe the CEP properties  of real liquids. 
Since the QGBSTM1 \cite{Bugaev_07_physrev} is also able to reproduce  the critical exponents \cite{Ivanytskyi} of  the same universality classes as the QGBSTM2, we may hope that  these models can be effectively used to 
describe the QCD endpoint properties   and they can help to experimentally  distinguish the 
CEP  case  from the  tricritical endpoint  case.

%
%\begin{table*}[!]
\begin{table}[t]

\begin{tabular}{|c|c|c|c|c|c|c|c|c|}
\hline
               &   \multicolumn{2}{c|}{2D Ising}   &         3D Ising                &
                            Simple                 &  \multicolumn{2}{c|}{O(4)}           \\
               &   \multicolumn{2}{c|}{Model}      &           Model                 &
                           Liquids                 &  \multicolumn{2}{c|}{Model}           \\ \hline
    $\xi^T$    &\multicolumn{2}{c|}{$\frac{8}{15}$}&         0.6010(1)               &
                           0.631(2)                &   \multicolumn{2}{c|}{0.55(1)}        \\ \hline
   $\zeta^+$   &     \multicolumn{2}{c|}{1}        &              1                  &
                              1                    &     \multicolumn{2}{c|}{1}            \\ \hline
   $\zeta^-$   &     $~~1~~$     &      $>1$       &              1                  &
                                 1                 &    $~~1~~$   &       $>1$             \\ \hline
    $\xi^+$    &     \multicolumn{2}{c|}{1}        &           1.0112(2)             &
                            0.937(23)              &   \multicolumn{2}{c|}{1.20(4)}        \\ \hline
    $\xi^-$    &     $~~1~~$     &      $>0$       &           0.9903(2)             &
                            0.910(25)              &    1.20(4)   &      $~~~>0~~~$        \\ \hline
   $\omega^*$  &      \multicolumn{2}{c|}{0}       &          $\emptyset$            &
                           $\emptyset$             &  \multicolumn{2}{c|}{0.19(6)}         \\ \hline
 $\omega^{**}$ &    \multicolumn{2}{c|}{$\ge0$}    &             $\ge0$              &
                             $\ge0$                &   \multicolumn{2}{c|}{$\ge0$}         \\ \hline
 $\omega^{***}$&     \multicolumn{2}{c|}{$\ge0$}   &             $\ge0$              &
                               $\ge0$              & \multicolumn{2}{c|}{$\ge0.19(6)$}     \\ \hline
\end{tabular}
\caption{The QGBSTM2 parameters which describe the corresponding exponents given in the Table \ref{table1}. The values of the parameter $\omega$ extracted from the expressions for $\alpha'$, $\alpha'_s$ and $\alpha'_c$ are marked with $^{*}$, $^{**}$ and $^{***}$, respectively. The symbol $\emptyset$ means that it is impossible to find the value of the corresponding parameter which allows us to describe the critical exponents shown in the Table \ref{table1}.}
\label{table3}
%
%\end{table*}
\end{table}
%

%%%%%%%%%%%%%%%%%%%%5 Conclusions %%%%%%%%%%%%%%%%%%%%%%%%%%%%%%%%%%%%%%%%%%%%%%%%%%%%%%%%

%
\section{Conclusions}
\label{secconclusions}

The practical necessity to describe   the thermodynamic functions of the QGP with the (tri)critical endpoint which has the required  properties stimulated the development of a variety of the exactly solvable statistical models \cite{Bugaev_07_physrev, FWM,Bugaev_09,  Goren:05,CGreiner:06, CGreiner:07, Koch:09, CGreiner:10}. 
Since it is not exactly know whether the QCD phase diagram endpoint is critical or tricritical it was necessary to developed the exactly solvable models for both cases. However, in contrast to  the tricritical endpoint case which was worked out  in many versions \cite{Bugaev_07_physrev, FWM,Bugaev_09,  Goren:05,CGreiner:06, CGreiner:07, Koch:09, CGreiner:10},  a formulation of a realistic exactly solvable model with the CEP required many years and took many additional efforts, because  neither the physical mechanism of its generation nor the mathematical properties of such a model were known before its development in \cite{Bugaev_09}. 
The main result of  \cite{Bugaev_09} clearly demonstrates  that 
 the traditional  framework of   liquid-gas PT models \cite{Fisher:Rev67,Fisher_67, Bondorf, Bugaev_00} has to be essentially modified   in order to  degenerate  the 1st order deconfinement PT into a CEP and into a cross-over. 
Furthermore, on the one hand the liquid-gas PT models require the vanishing of  surface tension coefficient at the CEP, and on the other hand  the positive values of the confining tube entropy unavoidably  demand for the negative surface tension coefficient  at the cross-over region \cite{String:09,Bugaev:12}.  The natural conclusion that follows from these findings is that 
at the cross-over region there should exist the curve of nil values of the surface tension coefficient which must
pass through the endpoint of the 1st order deconfinement PT.  If the nil line of the surface tension coefficient does  cross the deconfinement PT curve at the endpoint only, then, as shown in   \cite{Bugaev_07_physrev}, such an endpoint of the deconfinement PT is the tricritical one and this is the case of  QGBSTM1. If,  however, the nil curve
of the surface tension coefficient matches the deconfinement PT curve as described above (see Fig. 1 for details), then the model has the CEP and it corresponds to the  QGBSTM2 \cite{Bugaev_09}. 

The present paper is devoted to the analysis of the critical exponents
of the  QGBSTM2 which are  necessary to be studied before applying the present model to
 describe  either  the lattice QCD thermodynamics or various  experimental data. 
 Such an analysis allowed us to find some general restrictions on the model parameters and figure out 
the important relations between them.  
Also 
it is found that  the QGBSTM2 exponents essentially differ from that ones of the FDM, SMM and QGBSTM1 despite  many similarities between these models. Thus, in the FDM and the  SMM the critical exponents can be expressed in terms  of a few major input  parameters like the Fisher topological exponent $\tau$ and  the exponent $\varkappa$ which is usually related to a dimension, whereas the 
QGBSTM1  critical exponents also depend  on two recently introduced parameters $\xi^T$ and 
$\chi$  \cite{Ivanytskyi}.  In addition to all these parameters  the  QGBSTM2  depends  on  the  input exponents  $\xi^\pm$,  $\zeta^\pm$ and $\omega$, but in this work it is found that, in contrast to the FDM, SMM and QGBSTM1, 
the QGBSTM2 exponents do not depend on the input exponents  $\varkappa$, $\chi$ and $\tau$, although here we for the first time  found that the  present model formulation is valid   for $\tau>3$ only. 

Here we also showed that neither with  the traditionally defined index $\alpha'$ of (\ref{EqXVIII}) nor with the specially introduced one $\alpha'_s$ \cite{Fi:70} it is possible to fulfill  the Fisher and Griffiths  scaling  inequalities, while the Liberman one is always obeyed.  Such a situation is known for  the models 
\cite{Fi:70,Bugaev_00,Bugaev_07_physrev, FWM}, but in the case of present model the index $\alpha'_s$ does not in general  allow one to recover the scaling inequalities (\ref{F1}) and (\ref{G}) as well. 
However, in the present work  we proved  that for  a physically motivated definition of the  index $\alpha_c^\prime$ which 
corresponds to the maximal value of the index $\alpha'$ of two pure phases,
the QGBSTM2 exponents recover  the Fisher   scaling
inequality (\ref{F1}), while the  Griffiths inequality (\ref{G}) in its traditional form is fulfilled  only in a case, 
when all scaling laws (\ref{F1})--(\ref{L}) become the equalities. 
It is also  shown that for this  traditional scaling regime the QGBSTM2 exponents reproduce the values of  critical indices of the  2-dimensional Ising model, of the simple liquids,  of the 3-dimensional Ising model and of the O(4) symmetric spin model  and this evidences for the fact that the bags (or  physical clusters) used in the QGBSTM2 are the relevant degrees of freedom at the CEP  for  all studied systems despite quite a different physics of  their phases.
The present analysis demonstrates  that the negative values of the 
$\alpha'_c$ index can be achieved with the help of newly introduced exponent $\omega$ in 
(\ref{EqXXXXIV}), which, as we explicitly demonstrated,  plays a decisive role in the recovering  of scaling inequalities.  

Besides the  traditional scaling regime,
we found   that the QGBSTM2 exponents  have  the generalized scaling regime, for which 
the  Griffiths inequality should be generalized to Eq. (\ref{EqXXXXXII}). Also here we showed that for 
the QGBSTM2 critical indices the Liberman inequality can be similarly generalized  to a stronger  relation  (\ref{EqXXXXXIII}).  It is clear that a similar results should be obtained 
for the systems whose surface tension at the vicinity of the CEP can be reduced to the form of Eq. (\ref{EqIX}). The case of the asymmetric phase diagrams with other surface tension parameterization  should be analyzed both experimentally and theoretically. The example of the present model suggests that for the  asymmetric phase diagrams the Fisher scaling inequality can be recovered by the proper definition of the index $\alpha'$, while for such phase diagrams the Griffiths scaling  inequality in its usual form  can be broken down. 
This can occur for the linear temperature dependency of the surface tension coefficient in both pure phases since $\xi^+ = \xi^- =1$ which is not too restrictive as it is demonstrated by   the FDM and by  the exactly solvable model of surface deformations  \cite{Bugaev_05_Physrev}. 
The other important constraint on such  systems  is that their  surface tension coefficient should change its sign while crossing the PT line.  Clearly, such a property is rather unusual for the ordinary liquids and is more resembling the transition from a mixture of gases into a strongly interacting  plasma, which, as one can argue \cite{Hosek:91,Hosek:93,String:09,Bugaev_07_physrev},  is the case for the QGP.

Also we believe that for the asymmetric phase diagrams the question of the proper definition of the critical exponents and the scaling relations between them   should be  
thoroughly investigated both experimentally and theoretically. In particular, in many works the index 
$\alpha'$ is difficult to determine directly and, hence, it is usually found from the hyperscaling relation \cite{Stanley:71,Lavis:99} or even from one of the scaling equalities 
(\ref{F1}) or (\ref{G}), while, as it is suggested by  the above analysis,  in some cases one can get the  unexpected  results for the traditional scaling inequalities.

%%%KAB 

\medskip

{\bf Acknowledgments.}  The authors appreciate the valuable comments of  I. N. Mishustin and L. M. Satarov. 
A.I.I., K.A.B. and G.M.Z. acknowledge  the partial  support of the Program 'Fundamental Properties of 
Physical Systems under Extreme Conditions' launched by the Section of Physics and Astronomy of National Academy of  Sciences of Ukraine.
The work of  A.S.S.  was supported in part by the Russian
Foundation for Basic Research, Grant No. 11-02-01538-a.

%\medskip

%%%%%%%%%%%%%%% Appendices 

\section{Appendices}

\subsection{The isothermal compressibility}
\label{secappendixA}
In this appendix we give the details of the isothermal compressibility evaluation. 
As usual, one can simplify 
the isothermal compressibility as $K_T=\frac{1}{\rho}\frac{d\rho}{dp}\bigl|_T \equiv \frac{1}{\rho^2}\frac{\partial\rho}{\partial\mu}$. An explicit expression for the baryonic density of pure phases is as follows
\begin{equation}
\rho^\pm=T\left[\frac{\partial z_M}{\partial\mu}+\frac{A_\mu-\frac{\partial\Sigma^\pm}{\partial\mu}uI_{\tau-\varkappa}}
{1-\frac{\partial F_H}{\partial s}+uI_{\tau-1}}\right] \,,
\end{equation}
from which one directly obtains its $\mu$ derivative as 
{\begin{eqnarray}
\label{EqXXXXXVI}
\frac{\partial\rho^\pm}{\partial\mu}&=&
T\left[\frac{\partial^2z_M}{\partial\mu^2}+ f_1+f_2\right],
\end{eqnarray}
where the following notations are introduced 
\begin{eqnarray}
&f_1 =
\frac{1}{\left(1-\frac{\partial F_H}{\partial s}+uI_{\tau-1}\right)}
\left[ \frac{\partial A_\mu}{\partial\mu}-
\frac{\partial^2\Sigma^\pm}{\partial\mu^2}uI_{\tau-\varkappa}  - \frac{\partial\Sigma^\pm}{\partial\mu}\frac{\partial u}{\partial\mu}
\right. 
\nonumber \\
\label{EqXXXXXVIa}
& \times I_{\tau-\varkappa} + \left. 
\frac{\partial\Sigma^\pm}{\partial\mu}u
\left(\frac{\partial\Delta^\pm z}{\partial\mu}I_{\tau-\varkappa-1}+
\frac{\partial\Sigma^\pm}{\partial\mu}I_{\tau-2\varkappa}\right)
\right]\, , 
\end{eqnarray}
and
\begin{eqnarray}
f_2 &=&
\frac{A_\mu-\frac{\partial\Sigma^\pm}{\partial\mu}uI_{\tau-\varkappa}}
{\left(1-\frac{\partial F_H}{\partial s}+uI_{\tau-1}\right)^2}
\left[\frac{\partial^2F_H}{\partial s\partial\mu}-\frac{\partial u}{\partial\mu}I_{\tau-1}\right.
\nonumber \\
\label{EqXXXXXVIb}
&+&\left.
u\left(\frac{\partial\Delta^\pm z}{\partial\mu}I_{\tau-2}+
\frac{\partial\Sigma^\pm}{\partial\mu}I_{\tau-1-\varkappa}\right)\right].
\end{eqnarray}
All the regular terms in the expressions  (\ref{EqXXXXXVI}), (\ref{EqXXXXXVIa}) and (\ref{EqXXXXXVIb}) can be neglected, since the isothermal compressibility behavior  near the CEP is defined by  the singular ones. Hence, keeping only the  singular terms we get
\begin{eqnarray}
\label{EqXXXXXX}
&T^{-1}&\frac{\partial\rho^\pm}{\partial\mu} \simeq  
 \frac{
\frac{\partial\Sigma^\pm}{\partial\mu}u\left(\frac{\partial\Delta^\pm z}{\partial\mu}I_{\tau-\varkappa-1}+
\frac{\partial\Sigma^\pm}{\partial\mu}I_{\tau-2\varkappa}\right)}
{1-\frac{\partial F_H}{\partial s}+uI_{\tau-1}}  \nonumber
\\
&+&\frac{\left(A_\mu-\frac{\partial\Sigma^\pm}{\partial\mu}uI_{\tau-\varkappa}\right)
u\left(\frac{\partial\Delta^\pm z}{\partial\mu}I_{\tau-2}+
\frac{\partial\Sigma^\pm}{\partial\mu}I_{\tau-1-\varkappa}\right)}
{\left(1-\frac{\partial F_H}{\partial s}+uI_{\tau-1}\right)^2}   \nonumber
\\
&-& \frac{
\frac{\partial^2\Sigma^\pm}{\partial\mu^2}uI_{\tau-\varkappa}}{1-\frac{\partial F_H}{\partial s}+uI_{\tau-1}} \,   .
\end{eqnarray}
In order to calculate  the critical exponent $\gamma'$ one, according to the suggestion of  \cite{Fi:70},  has to evaluate 
 the difference of  the isothermal compressibilities  across the PT line $\Delta K_T=\left[\frac{1}{\rho^+}\frac{\partial\rho^+}{\partial\mu}-
\frac{1}{\rho^-}\frac{\partial\rho^-}{\partial\mu}\right]_{T=T_c}$. 
Requiring that such a  quantity remains finite everywhere at the phase coexistence curve,  except for the  CEP,  one has to demand that all the integrals $I_{\tau-q}$ in Eq. (\ref{EqXXXXXX}) are  finite  for $T=T_c$, including  $T =T_{cep}$.
 The analysis shows that  it is sufficient to require that all such integrals, including  $I_{\tau-2}$, converge for $\tau>3$  at the PT line,  where $\Delta z=0$ and $\Sigma^\pm=0$. Then, the only possible  singularity  at the CEP   in the  expression for $\frac{\partial\rho^\pm}{\partial\mu}$ is provided by  the term which is  proportional to $\frac{\partial^2\Sigma^\pm}{\partial\mu^2}$.  Then for  the  isothermal compressibility one finds 
\begin{eqnarray}
\label{EqXXXXXXI}
K_T^\pm\simeq-(\rho^\pm)^{-2}\frac{\partial^2\Sigma^\pm}{\partial\mu^2}\cdot
\frac{TuI_{\tau-\varkappa}}{1-\frac{\partial F_H}{\partial s}+uI_{\tau-1}}.
\end{eqnarray}
Just this expression is used to evaluate  the  right hand side of  Eq. (\ref{EqXXIX}). 

%%%KAB

\subsection{Volume fraction of pure phases}
\label{secappendixB}

In order to establish the relation between the indices $\beta$ and $\chi$ it is convenient to
 analyze the behavior of the pure phase volume fractions at the critical isochore near the  CEP.
The QGBSTM2  critical isochore completely lies  inside the mixed phase.   The critical  baryonic density can be described by the volume fraction of hadrons $\lambda$ (the  QGP volume fraction  is $1-\lambda$, respectively) as
\begin{equation}
\rho_{cep}=\lambda\rho^-|_{T=T_c}+(1-\lambda)\rho^+|_{T=T_c},
\end{equation}
where $\rho^-$ and $\rho^+$ are, respectively, the baryonic densities of hadronic phase and QGP. From the previous equation one finds 
\begin{equation}
\lambda=\frac{\rho^+-\rho_{cep}}{\rho^+-\rho^-} \quad {\rm and}\quad 1-\lambda=\frac{\rho_{cep}-\rho^-}{\rho^+-\rho^-}.
\end{equation}
Using an evident relation $\rho_{cep}=\rho_M|_{cep}$ and the definition of the baryonic density  one obtains
\begin{eqnarray}\label{Eq6X}
&&(\rho^\pm-\rho_{cep})_{T=T_c}=(T_c-T_{cep})\frac{\partial z_M}{\partial\mu}\biggl|_{cep}
\nonumber \\
&&+T_c\left[\frac{A_\mu-\frac{\partial\Sigma^\pm}{\partial\mu}uI_{\tau-\varkappa}}
{1-\frac{\partial F_H}{\partial z}+uI_{\tau-1}}+
\frac{\partial z_M}{\partial\mu}-\frac{\partial z_M}{\partial\mu}\biggl|_{cep}\right]_{T=T_c}
\hspace*{-0.2cm}.\hspace*{0.6cm}
\end{eqnarray}
Applying 
the   parameterization  defined by Eqs. (\ref{EqIX}), (\ref{EqXIII}) and (\ref{EqXVII}) to the right hand side of (\ref{Eq6X}),
one can easily  demonstrate that either $(\rho^\pm-\rho_{cep})_{T=T_c}\sim t^{\min(\chi,\beta^\pm,1,\frac{1}{\xi^T})}$ for $\zeta^\pm=1$ or  $(\rho^\pm-\rho_{cep})_{T=T_c}\sim t^{\min(\chi,1,\frac{1}{\xi^T})}$ for $\zeta^\pm>1$. Note that for this proof it is necessary to expand the difference $\frac{\partial z_M}{\partial\mu}-\frac{\partial z_M}{\partial\mu}\bigl|_{cep}$  up to linear terms of  $\mu-\mu_{cet}$ and $T_c-T_{cep}$. Finally, using the definition of the index $\beta$, i.e. $(\rho^+-\rho^-)_{T=T_c}\sim t^\beta$,  and the feature of  the present model $\zeta^+ =1$
  we find the following behavior  of the volume fractions
\begin{eqnarray}
\lambda&\sim& t^{\min(\chi,\beta^+,1,\frac{1}{\xi^T})-\beta},\\
1-\lambda&\sim&\left\{\hspace*{-0.1cm}\begin{array}{lr}
 t^{\min(\chi,\beta^-,1,\frac{1}{\xi^T})-\beta},  &\hspace*{-0.1cm}
{\rm for}~ \zeta^-=1 \,,\\
 t^{\min(\chi,1,\frac{1}{\xi^T})-\beta},  &\hspace*{-0.1cm}
{\rm for}~ \zeta^->1 \,.
\end{array}\right.
\end{eqnarray}
in the vicinity of the CEP. 
Combining  these results with the fact that both $\lambda$ and $(1-\lambda)$ are finite at the CEP, we obtain a very important consequences that
$\min(\chi,\beta^+,1,\frac{1}{\xi^T})-\beta\ge0$,  and either  $\min(\chi,\beta^-,1,\frac{1}{\xi^T})_{\zeta^-=1}-\beta\ge0$ for  $\zeta^-=1$ or $\min(\chi,1,\frac{1}{\xi^T})_{\zeta^->1}-\beta\ge0$ for  $\zeta^->1$. An explicit expression for the index $\beta$ (\ref{beta}) allows us to rewrite all these conditions as 
\begin{equation}
\beta\le\min\left(\chi,1,\frac{1}{\xi^T}\right),
\end{equation}
which holds always. From the inequality above one immediately deduces that $\beta \le \chi$.
These results are important for finding the index $\alpha'_c$.  Indeed, recalling the expression (\ref{beta}) one can write 
\begin{equation}
\min\left[ \min \left(\chi,\beta^+,1,\frac{1}{\xi^T} \right), ~ \min\left(\chi,\beta^-,1,\frac{1}{\xi^T} \right)   \right] = \beta\,,
\end{equation}
for the case $\zeta^-=\zeta^+=1$ and 
\begin{equation}
\min\left[ \min \left(\chi,\beta^+,1,\frac{1}{\xi^T} \right), ~ \min\left(\chi,1,\frac{1}{\xi^T} \right)   \right] = \beta\,,
\end{equation}
for the case $\zeta^->\zeta^+=1$.
From the above equations one can immediately deduce that $\max( (\rho^+-\rho_M),~ (\rho^--\rho_M) ) \sim t^\beta$ at the vicinity of the CEP, since $\beta \ge 0$.  Thus, here we showed that  the term staying  inside  the square brackets on the right hand side of    Eq. (\ref{EqXXXXIII}), indeed, behaves as 
$t^{\frac{1}{\xi^T}+\beta-2}$. 

%

%%%%%%%%%%%%%%%%%%%%  The bibliography  %%%%%%%%%%%%%%%%%%%%%%%%%%%%%%%%%%%%%%%%%%%%%%%%%

\end{document}